\documentclass[usenatbib]{mnras}
\usepackage{lmodern}

\usepackage[T1]{fontenc}
\usepackage[utf8]{inputenc}
\usepackage{color}
\usepackage{subfigure}
\usepackage{amsmath}
\usepackage{amssymb}
\usepackage{graphicx}
\graphicspath{{figures/}}
\usepackage{esint}
\usepackage{multirow}
\usepackage{cuted}
\usepackage{makecell}
\usepackage{grffile}
\usepackage[normalem]{ulem}
\AtBeginDocument{\hypersetup{
linkcolor=linkblue,
citecolor=linkorange,
urlcolor=urlblue}}

\providecommand{\eprint}[1]{\href{http://arxiv.org/abs/#1}{#1}}
\providecommand{\adsurl}[1]{\href{#1}{ADS}}

\usepackage{ifthen}\def\eprinttmp@#1arXiv:#2 [#3]#4@{\ifthenelse{\equal{#3}{x}}{\href{http://arxiv.org/abs/#1}{#1}}{\href{http://arxiv.org/abs/#2}{arXiv:#2} [#3]}}
\renewcommand{\eprint}[1]{\eprinttmp@#1arXiv: [x]@}

\definecolor{urlblue}{rgb}{0,0,0.9}
\definecolor{linkblue}{rgb}{0,0,.8}
\definecolor{linkgreen}{rgb}{0,0.45,0}
\definecolor{linkpurple}{rgb}{0.7,0.0,0.4}
\definecolor{linkorange}{rgb}{0.7,0.1,0.0}

\makeatletter

\pdfpageheight\paperheight
\pdfpagewidth\paperwidth

\definecolor{somegreen}{cmyk}{0,0.49,0.98,0.09}
\definecolor{red}{rgb}{1,0,0}
\definecolor{magenta}{cmyk}{0,1,0,0}
\definecolor{lavender}{cmyk}{0.16,0.67,0,0.57}
\definecolor{darkgreen}{rgb}{0,0.65,0.05}
\definecolor{antiquefuchsia}{rgb}{0.33, 0.1, 0.89}

\let\jnl@style=\rm
\def\ref@jnl#1{{\jnl@style#1}}

\def\aj{\ref@jnl{AJ}}                   % Astronomical Journal
\def\actaa{\ref@jnl{Acta Astron.}}      % Acta Astronomica
\def\araa{\ref@jnl{ARA\&A}}             % Annual Review of Astron and Astrophys
\def\apj{\ref@jnl{ApJ}}                 % Astrophysical Journal
\def\apjl{\ref@jnl{ApJ}}                % Astrophysical Journal, Letters
\def\apjs{\ref@jnl{ApJS}}               % Astrophysical Journal, Supplement
\def\ao{\ref@jnl{Appl.~Opt.}}           % Applied Optics
\def\apss{\ref@jnl{Ap\&SS}}             % Astrophysics and Space Science
\def\aap{\ref@jnl{A\&A}}                % Astronomy and Astrophysics
\def\aapr{\ref@jnl{A\&A~Rev.}}          % Astronomy and Astrophysics Reviews
\def\aaps{\ref@jnl{A\&AS}}              % Astronomy and Astrophysics, Supplement
\def\azh{\ref@jnl{AZh}}                 % Astronomicheskii Zhurnal
\def\baas{\ref@jnl{BAAS}}               % Bulletin of the AAS
\def\bac{\ref@jnl{Bull. astr. Inst. Czechosl.}}
\def\caa{\ref@jnl{Chinese Astron. Astrophys.}}
\def\cjaa{\ref@jnl{Chinese J. Astron. Astrophys.}}
\def\icarus{\ref@jnl{Icarus}}           % Icarus
\def\jcap{\ref@jnl{J. Cosmology Astropart. Phys.}}
\def\jrasc{\ref@jnl{JRASC}}             % Journal of the RAS of Canada
\def\memras{\ref@jnl{MmRAS}}            % Memoirs of the RAS
\def\mnras{\ref@jnl{MNRAS}}             % Monthly Notices of the RAS
\def\na{\ref@jnl{New A}}                % New Astronomy
\def\nar{\ref@jnl{New A Rev.}}          % New Astronomy Review
\def\pra{\ref@jnl{Phys.~Rev.~A}}        % Physical Review A: General Physics
\def\prb{\ref@jnl{Phys.~Rev.~B}}        % Physical Review B: Solid State
\def\prc{\ref@jnl{Phys.~Rev.~C}}        % Physical Review C
\def\prd{\ref@jnl{Phys.~Rev.~D}}        % Physical Review D
\def\pre{\ref@jnl{Phys.~Rev.~E}}        % Physical Review E
\def\prl{\ref@jnl{Phys.~Rev.~Lett.}}    % Physical Review Letters
\def\pasa{\ref@jnl{PASA}}               % Publications of the Astron. Soc. of Australia
\def\pasp{\ref@jnl{PASP}}               % Publications of the ASP
\def\pasj{\ref@jnl{PASJ}}               % Publications of the ASJ
\def\rmxaa{\ref@jnl{Rev. Mexicana Astron. Astrofis.}}%
\def\qjras{\ref@jnl{QJRAS}}             % Quarterly Journal of the RAS
\def\skytel{\ref@jnl{S\&T}}             % Sky and Telescope
\def\solphys{\ref@jnl{Sol.~Phys.}}      % Solar Physics
\def\sovast{\ref@jnl{Soviet~Ast.}}      % Soviet Astronomy
\def\ssr{\ref@jnl{Space~Sci.~Rev.}}     % Space Science Reviews
\def\zap{\ref@jnl{ZAp}}                 % Zeitschrift fuer Astrophysik
\def\nat{\ref@jnl{Nature}}              % Nature
\def\iaucirc{\ref@jnl{IAU~Circ.}}       % IAU Cirulars
\def\aplett{\ref@jnl{Astrophys.~Lett.}} % Astrophysics Letters
\def\apspr{\ref@jnl{Astrophys.~Space~Phys.~Res.}}
\def\bain{\ref@jnl{Bull.~Astron.~Inst.~Netherlands}}
\def\fcp{\ref@jnl{Fund.~Cosmic~Phys.}}  % Fundamental Cosmic Physics
\def\gca{\ref@jnl{Geochim.~Cosmochim.~Acta}}   % Geochimica Cosmochimica Acta
\def\grl{\ref@jnl{Geophys.~Res.~Lett.}} % Geophysics Research Letters
\def\jcp{\ref@jnl{J.~Chem.~Phys.}}      % Journal of Chemical Physics
\def\jgr{\ref@jnl{J.~Geophys.~Res.}}    % Journal of Geophysics Research
\def\jqsrt{\ref@jnl{J.~Quant.~Spec.~Radiat.~Transf.}}
\def\memsai{\ref@jnl{Mem.~Soc.~Astron.~Italiana}}
\def\nphysa{\ref@jnl{Nucl.~Phys.~A}}   % Nuclear Physics A
\def\physrep{\ref@jnl{Phys.~Rep.}}   % Physics Reports
\def\physscr{\ref@jnl{Phys.~Scr}}   % Physica Scripta
\def\planss{\ref@jnl{Planet.~Space~Sci.}}   % Planetary Space Science
\def\procspie{\ref@jnl{Proc.~SPIE}}   % Proceedings of the SPIE

\newcommand{\dd}{\textrm{d}}

\newcommand{\one}{$1\times 2$pt}
\newcommand{\three}{$3\times 2$pt}
\newcommand{\six}{\ensuremath{6\times 2}pt}

\makeatother

\title[The lure of sirens: distance + velocities]{The lure of sirens: joint distance and velocity measurements with third generation detectors}

\author[Alfradique, Quartin, Amendola, Castro \& Toubiana]{Viviane Alfradique$^{1}$, Miguel Quartin$^{1,2,3}$, Luca Amendola$^{3}$, Tiago Castro$^{4, 5, 6}$ \newauthor and Alexandre Toubiana$^{7}$ \\
$^{1}$Instituto de Física, Universidade Federal do Rio de Janeiro,
21941-972, Rio de Janeiro, RJ, Brazil\\
$^{2}$Observatório do Valongo, Universidade Federal do Rio de Janeiro,
20080-090, Rio de Janeiro, RJ, Brazil\\
$^{3}$Institute of Theoretical Physics, Heidelberg University, Philosophenweg 16, 69120 Heidelberg, Germany\\
$^{4}$INAF -- Osservatorio Astronomico di Trieste, 34131, Trieste, Italy \\
$^{5}$IFPU -- Institute for Fundamental Physics of the Universe, 34151, Trieste, Italy\\
$^{6}$INFN -- Sezione di Trieste, 34100, Trieste, Italy \\
$^{7}$Max Planck Institute for Gravitational Physics (Albert Einstein Institute), Am Mühlenberg 1, Potsdam-Golm, 14476, Germany
}

\date{\today }

\begin{document}
\label{firstpage}
\pagerange{\pageref{firstpage}--\pageref{lastpage}}

\maketitle

\begin{abstract}
   The next generation of detectors will detect gravitational waves from binary neutron stars at cosmological distances, for which around a thousand electromagnetic follow-ups may be observed per year. So far, most work devoted to the expected cosmological impact of these standard sirens employed them only as distance indicators. Only recently their use as tracers of clustering, similar to what already proposed for supernovae, has been studied. Focusing on the expected specifications of the Einstein Telescope (ET), we forecast here the performance on cosmological parameters of future standard sirens as both distance and density indicators, with emphasis on the linear perturbation growth index and on spatial curvature.  We improve upon previous studies in a number of ways: a more detailed analysis of available telescope time, the inclusion of more cosmological and nuisance parameters, the Alcock-Paczynski correction, the use of sirens also as both velocity and density tracers, and a more accurate estimation of the distance posterior. We find that the  analysis of the clustering of sirens improves the constraints on $H_0$ by 30\% and on $\Omega_{k0}$ by over an order of magnitude, with respect to their use merely as distance indicators. With 5 years of joint ET and Rubin Observatory follow-ups we could reach precision of 0.1 km/s/Mpc in $H_0$ and $0.02$ in $\Omega_{k0}$ using only data in the range $0<z<0.5$.   We also find that the use of sirens as tracers of density, and not only velocity, yields good improvements on the growth of structure constraints.
\end{abstract}


\maketitle

\section{Introduction}

Since the breakthrough of gravitational wave (GW) astronomy with the first direct detection in 2015~\citep{Abbott:2016blz} by the LIGO/Virgo collaboration (LVC)~\citep{TheLIGOScientific:2014jea,TheVirgo:2014hva}, several events have been reported. The total number is changing fast, and currently approaches one hundred  detections~\citep{LIGOScientific:2018mvr,LIGOScientific:2020ibl,LIGOScientific:2021djp}. Among those, of special importance was the binary neutron star (BNS) GW170817~\citep{TheLIGOScientific:2017qsa}, which was accompanied by an electromagnetic (EM) counterpart~\citep{GBM:2017lvd}. This breakthrough \emph{multimessenger} observation provided a joint precise determination of both the distance and redshift of the event, allowing its use for the first time as standard siren, and in particular to measure the Hubble constant $H_0$ independently~\citep{LIGOScientific:2017adf}.

However,  the constraint on the luminosity distance $d_L$ of the single event GW170817 translates into a large uncertainty on $H_0$.
Better constraints will be achieved with additional GW detections, but only one additional EM counterpart candidate has been reported so far, to wit for the binary black hole (BBH) coalescence GW190521~\citep{LIGOScientific:2020ibl,Graham:2020gwr}, and moreover this association is still uncertain \citep{Ashton:2020kyr,Bustillo:2021tga}. Though this slightly improves the error on $H_0$~\citep{Mukherjee:2020kki,Chen:2020gek,Bustillo:2021tga}, the latter remains much larger than the one  obtained indirectly with the cosmological microwave background (CMB)~\citep{Planck:2018vyg} or clustering~\citep{Philcox:2020vvt}, or  directly with Cepheids~\citep{Riess:2021jrx} measurements and the Tip of the Red Giant Branch methods~\citep{Freedman:2019jwv}.

The need for strong and independent constraints on $H_0$ nevertheless cannot be overestimated. As well-known, the long standing discrepancy between some (mostly low-$z$) estimates of $H_0$ and other (mostly high-$z$) estimates, did not disappear, and in fact intensified, with more and more precise measurements \citep[for a recent review, see][]{Abdalla:2022yfr}. This tension might well be the first indication of a failure of the standard cosmological model, and deserves the most intense scrutiny. Moreover, with future upgrades and observing runs of the LVC instruments there are great expectations that standard sirens might clarify the $H_0$ conundrum.

Other methods to measure cosmological parameters with GWs have been proposed which preclude the need of an electromagnetic counterpart~\citep{LIGOScientific:2021aug}. For instance, one can: (i) use the existence of sharp features in the mass distribution of BBHs such as a characteristic mass scale or a mass-gap~\citep{Chernoff:1993th,Taylor:2011fs,Farr:2019twy,Leyde:2022orh,Ezquiaga:2022zkx}; (ii) perform statistical host identification through correlation with galaxy catalogues in order to constrain the GW redshifts~\citep{Schutz:1986gp,DelPozzo:2011vcw,Chen:2017rfc,Finke:2021aom,Mukherjee:2022afz}; (iii) break the mass-redshift degeneracy if the tidal deformation can be measured in the waveform~\citep{Messenger:2011gi}; (iv) analyse jointly the spatial clustering of GW (in luminosity-distance space) and galaxies (in redshift space)~\citep{Mukherjee:2020hyn,Diaz:2021pem};
(v) rely also purely on the redshift distribution of sources in order to constrain cosmology~\citep{Leandro:2021qlc}, if thousands of BBHs are available. But it is still unclear how competitive these methods will be in the future when compared to using BNS mergers, and we will not consider these possibilities in this work.

In the next decade, third generation GW detectors, such as the Einstein Telescope (ET) \citep{Hild:2010id,Punturo:2010zz,Ballmer:2015mvn} and Cosmic Explorer (CE) \citep{LIGOScientific:2016wof,Reitze:2019iox}, will detect BNS mergers at high redshifts and most BBH coalescences in the Universe with total masses $M \sim 2-2000 M_\odot$~\citep{Maggiore:2019uih}. This will greatly increase the number of candidates for multimessenger observations, allowing us to perform precise measurement of cosmological parameters.

Traditionally, standard candles such as type Ia supernovae~(SN) and standard sirens have been primarily considered as tools  to build a Hubble-Lemaître diagram, and thus to constrain the background expansion of the universe. More recently, it has been realized that standard candles can also be powerful tools to measure cosmological perturbation parameters, which model the amount of structure in the universe and its growth. For higher redshift sources, \cite{Quartin:2013moa} showed that the non-Gaussianities in the Hubble diagram residuals introduced by weak-lensing can yield precise constraints with the upcoming LSST survey. For low and intermediate redshifts, \cite{Gordon:2007zw} proposed instead the measurements of the correlations between the supernova Hubble residual points as a probe of the peculiar velocity field. Both techniques have since been refined~\citep{Johnson:2014kaa,Macaulay:2016uwy,Howlett:2017asw}, and low-precision constraints from real data already been established~\citep{Castro:2015rrx,Qin:2019axr,DES:2020kbf}.

Since the energy in EM waves obeys the inverse-square law but GW amplitudes decay instead  with a single power of the distance, in relative terms it is simpler to detect high-redshift BNS GWs than to observe their EM counter-parts. It is thus expected that standard sirens observed with the help of third-generation GW facilities will have higher completeness at lower redshifts. This is precisely the range in which peculiar velocity effects are more relevant, and therefore in this work we will neglect lensing and instead focus on the peculiar velocity measurements made possible by this next generation of GW observatories.

\cite{Palmese:2020kxn} made the first forecasts for this science case. In this work we will revisit this idea improving it in a number of ways. In particular, we will: (i) make more detailed estimates of the amount of telescope time it takes to perform the EM follow-ups with different current and upcoming telescopes following~\cite{Chen:2020zoq}; (ii) investigate the benefits of the \six\ method  recently proposed by~\cite{Quartin:2021dmr}, which makes use of the standard sirens as both velocity and density tracers; (iii) include a larger number of cosmological and nuisance parameters to account for both linear bias and non-linear redshift-space distortions (RSD); (iv) include the Alcock-Paczynski (AP) corrections~\citep{Alcock:1979mp}; (v) make joint forecasts for clustering and traditional distance measurements; (vi) perform a more accurate determination of the distance posterior by adopting the analytical expressions in~\cite{Chassande-Mottin:2019nnz}.

As discussed in \cite{Torres-Orjuela:2018ejx,Torres-Orjuela:2020dhw}, the analogue of the ``beaming effect'' for GWs, due to the peculiar velocity of the source relative to the observer, affects the response the response of the detector to the $+$ and $\times$ polarizations in different manners and in addition leads to a mixing between the harmonics of the GW signal. This translates into a phase shift, which would potentially be detectable and would allow to break the degeneracy between source-frame mass and redshift and to measure the peculiar velocity of the source. However, for low peculiar velocities and nearly equal mass ratio systems as we are interested here, this effect should be subdominant and we will therefore neglect it in what follows.

This paper is divided as follows. In Section~\ref{sec:gw-catalog} we discuss the construction of our simulated GW catalogs. This is followed by calculations of the required amount of telescope time in order to perform the EM follow-ups, in Section~\ref{sec:EM-obs}. The final cosmological forecasts are presented in Section~\ref{sec:forecasts}, and the conclusions are discussed in Section~\ref{sec:discussions}. Details on the distance posterior approximations, on the effect of simultaneous BNS mergers, and on how the results depend on the distance uncertainties are covered in three appendices.

\section{Gravitational wave observations}\label{sec:gw-catalog}

\subsection{Catalogue generation} \label{sec:fiducial}

In this work we will focus on GWs observations accompanied with EM counterparts, that is called ``bright sirens''. For simplicity, we will consider that the bright sirens are generated only by the BNS mergers, ignoring the contribution of NS-BH binaries since the estimate of its merger rate is much lower than that found for BNS \citep{LIGOScientific:2021psn}. In order to construct a mock catalog of the BNS mergers observed by ET, we first assume that the BNS merger population is distributed within the cosmological volume through the redshift distribution $p\left(z\right)$ that is written as a function of the merger rate per redshift, in the observer frame $\mathcal{R}\left(z\right)$ ($\equiv \frac{dN}{dt_{\rm obs}dz}$, i.e. is the number of mergers per redshift and observer-frame time):
\begin{equation}
		p\left(z\right) = A \mathcal{R}\left(z\right) = A \frac{\dd V_{c}}{\dd z} \frac{\mathcal{R}_{m}\left(z\right)}{1+z},
	\end{equation}
where $A$ is a normalization constant ensuring that the integration of $p\left(z\right)$ goes to unity over $0<z<z_{\rm max}$, $dV/dz$ is the comoving volume element, and $\mathcal{R}_{m}\left(z\right)$ is the merger rate per comoving volume in the source frame. We assume that $\mathcal{R}_{m}$ follow the Madau-Dickinson star formation rate~\citep{Madau:2014bja}:
\begin{equation}
		\mathcal{R}_{m}\left(z\right) = \mathcal{R}_{0}\frac{1.00257(1+z)^{2.7}}{(1+\left(1+z/2.9\right)^{5.6}},
\end{equation}
with a local volumetric rate of $\mathcal{R}_{0}=300\,\rm{Gpc}^{-3}\rm{yr}^{-1}$, that agrees with the latest results of the LVC~\citep{LIGOScientific:2021psn}.\footnote{This merger rate is however still poorly constrained if one accounts for model systematics. Its 95\% confidence interval is very broad: $10-1700$ $\rm{Gpc}^{-3}$ yr$^{-1}$.} We neglect for simplicity the still very uncertain time delay between star formation and merger, although some authors consider it as a stochastic Poisson variable with a time scale of many Gyr~\citep{Safarzadeh:2019pis,deSouza:2021xtg}. To reduce computational time, we truncate the volume integral at $z=5$, which is enough to cover the ET detection range as we will show.

We draw NS masses $m_1$ and $m_2$ from a Gaussian distribution with mean $1.4\,\rm{M}_{\sun}$ and standard deviation $0.2\,\rm{M}_{\sun}$, restricted to the range 1–3 $\rm{M}_{\sun}$, which is in agreement with the limit inferred by the third observing run of LIGO–Virgo \citep[see section V.B of][]{LIGOScientific:2021psn}. Due to propagation in a Universe described by the Friedmann–Lemaître–Robertson–Walker metric, the observed masses in the detector frame are redshifted relative to the source-frame ones: $m_{1,2} \rightarrow m_{1,2}(1+z)$. We recall that the chirp mass is defined as $\mathcal{M}=(m_1^3m_2^3/(m_1+m_2))^{1/5}$, this is the parameter that drives the evolution of the binary at leading post-Newtonian order \citep{Blanchet:2013haa}. It is redshifted in the same way as the individual masses. Spins are isotropically oriented assuming uniform priors for both $s_{z1}$ and $s_{z2}$ in the range $[-0.05, \,0.05]$. We then take their projection on the orbital angular momentum axis and discard the transverse components. Tidal deformabilites are distributed according to a uniform distribution $U(0,5000)$. The events are assumed to be isotropically distributed and randomly oriented. Finally, the coalescence phase ($\varphi_c$) and the polarization angle ($\psi$) are both distributed according to a uniform distribution $U(0,2\pi)$.

The total number of BNS mergers \textit{N} can be simply computed using the definition $N = \Delta_{\rm obs}t_{\rm obs}\mathcal{R}_{\rm max}$, where $\mathcal{R}_{\rm max}=\int_{0}^{z_{\rm max}}\mathcal{R}\left(z\right)dz$, $\Delta_{\rm obs}$ is the duty cycle, and $t_{\rm obs}$ is the observation time. Assuming the flat $\Lambda$CDM model with $\Omega_{m0}=0.3$ and $H_{0}\equiv 100\,h$ km/s/Mpc $= 70\,$km/s/Mpc as our fiducial cosmological model, we find that the total number of BNS mergers is $N\approx 8\times 10^{5}$ per year of observation, assuming $\Delta_{\rm obs}=0.8$ for the ET~\citep{Belgacem2019}.

\subsection{Detection}

Here we will consider the prospects for detecting BNS mergers with third generation GW detectors, which are expected to take over Advanced Virgo in the 2030's. We will use the Europe-based ET specifications as our baseline third generation configuration, but similar results are expected for the USA-based CE. Both the number of events and the GW parameter estimation could be further improved by combining both detectors, or a combination of second and third generation facilities, but we will conservatively focus only on ET forecasts here.

ET consists of three Michelson interferometers arranged in a triangular shape, with 10-km-long arms and 60$^{\circ}$ opening angle, and can be seen as a combination of three noise-uncorrelated detectors. ET is sensitive to GWs in the $1-10^{4}$ Hz band, with a level of noise an order of magnitude lower than current detectors. This improvement will be achieved thanks to the longer arms, but also to the use of cryogenic technologies to reduce thermal noise and quantum technology to reduce the high-frequency quantum shot noise. As a result, ET will observe BNS mergers for much longer, up to tens of hours, improving on the parameter estimation. Moreover, combining the three noise-uncorrelated detectors will allow us to triangularise the signal, providing an accurate sky localization~\citep{Mills:2017urp,Chan:2018csa}.

For each BNS merger in our mock catalogue, we simulate its GW signal using the frequency domain approximant IMRPhenomD$\_$NRTidal, which uses the work of~\cite{Dietrich:2017aum} to include a 5PN modification to the phase due to tidal effects \citep{Damour:1984rbx,Flanagan:2007ix,Wade:2014vqa}.
The signal in each ET detector is given by a combination of the two GW polarisations, weighted by the antenna pattern functions \citep{1987MNRAS.224..131S,1987MNRAS.226..829T}:
\begin{equation}
 h_i=F_i^{+}h_{+}+F_i^{\times}h_{\times}.
\end{equation}
The expressions for $F_i^{+,\times}$ can be found in \cite{Regimbau:2012ir}. The total network signal-to-noise-ratio (SNR) is defined as
\begin{equation}
    \rho_{\rm net} = \left[\sum_{i=1}^{3}\rho_{i}^{2}\right]^{1/2} \;,
\end{equation}
where
\begin{equation}
    \rho_{i}^{2} = 4\int_{0}^{\infty}\frac{\left|h_i\left(f\right)\right|^{2}}{S_{n}\left(f\right)}\dd f \,
\end{equation}
is the SNR in the $i$-th ET detector. The power spectral density, $S_n(f)$, measures the level of noise (assumed to be stationary and Gaussian) in the detector at each frequency. We consider an event to be detectable if $\rho_{\rm net}\geq12$. We compute SNRs using the Bilby package~\citep{Ashton:2018jfp}.

\subsection{Parameter estimation}\label{sec:gw-params}

The parameters of a source are estimated using Bayes' theorem to obtain their posterior distribution from observed data. For our analysis, we are mostly interested in the precision of the sky localization for each event, in order to determine if there should be an EM follow-up, as well as the measurement error on the luminosity distance, which will be combined with EM measurement of redshift to infer cosmological parameters. Therefore, instead of performing computationally expensive Bayesian analyses, we use the analytic approach of \cite{Cutler:1994ys,Chassande-Mottin:2019nnz} to estimate the error on $d_L$. Assuming the sky localization is known from an EM counterpart and that intrinsic parameters (masses and spins) are not strongly correlated with inclination and distance, it provides the posterior distribution for the latter marginalised over polarization and phase. As illustrated in \cite{Chassande-Mottin:2019nnz}, this is a very good approximation in most cases, except for very few events where the GW polarization becomes degenerate.
These expressions are very convenient to perform forecasts, while at the same time being more accurate than simpler fits which relied exclusively on the relative distance errors of a given GW event~\citep{Zhao:2010sz}. In Appendix~\ref{app:app} we review BNS distance estimations and illustrate the performance of these expressions for the case of the ET.

For the sky localization, we assume $\Delta \Omega \propto 1/\rho_{\rm net}^2$, using the full Bayesian results for observations of BNSs with LIGO/Virgo network~\citep{DelPozzo:2018dpu} to calibrate the proportionality constant. Since triangularisation can be performed using the three ET detectors, we expect such a scaling with the SNR should hold.

\subsection{Comparisons between standard siren and supernova distances}\label{sec:kn-sn}

It is interesting to compare the capabilities of both supernovae (SN) and BNS mergers with EM counterparts as standard candles. The former has a much higher expected rate, to wit  $\mathtt{r}_{Ia} = 21000 \, (1+z)^{1.95}$/(yr Gpc${}^3$)~\citep{Cappellaro:2015,Amendola:2019lvy}. On the other hand, by themselves SN cannot measure $H_0$ due to their unknown absolute magnitudes $M_B$, and instead rely on calibration methods with external data such as Cepheids. Moreover, SN surveys indicate the presence of an intrinsic scatter which in the best cases, with infrared observations, is around $0.1\,$mag~\citep{Avelino2019}. BNS sirens on the other hand, as far as currently known, have precision limited only by the GW SNR. This means that as GW detectors improve, we should find some high SNR events with distances which are more precise than those of SN. Moreover,  both the large number of already identified possible sources of systematic effects on SN~\citep[see, e.g.,][]{Howell:2010vd} and in general the empirical nature of the SN light-curve fitting mean that it will be difficult to keep systematic effects subdominant in upcoming SN data.

In Section~\ref{sec:forecasts} below we make quantitative comparisons between BNS and SN as cosmological probes when combining GW and EM measurements, under the assumption of no relevant systematic sources of uncertainties in either one.

\section{Electromagnetic observations}\label{sec:EM-obs}

BNS coalescences are expected to originate various types of EM counterparts across the spectrum. The most promising counterparts for obtaining redshifts for the number of BNS mergers expected to be detected by the ET are the kilonovae (KN). These are optical/near-IR emissions caused by the decay of heavy ions via the r-process, which occurs in the mass ejecta of the BNS merger~\citep{Li:1998bw,Metzger:2010sy,Barnes:2013wka}. This emission can be viewed up to days-weeks after their production. Besides the optical/near-IR radiation, the ejected KN can still interact with the interstellar medium, producing a forward shock that emits  radiation in the radio waveband.
Another expected counterpart are GRBs. However, those emissions are highly beamed along the binary's orbital axis. As discussed in~\cite{Chen:2020zoq}, this high collimation together with the typical brightness of these GRBs mean that the observable number of these multimessenger events are considerably smaller than those involving the kilonovae. We will therefore for simplicity only consider the latter in this work. For this reason, we will also henceforth use the term kilonova as a shorthand for any BNS mergers with EM counterparts, and note that the inclusion of other EM counterparts would only improve the precision of our forecasts.

The observation of an EM counterpart may provide an additional constraint on the inclination angle, which helps to break the typical degeneracy with the luminosity distance, reducing in turn the uncertainty on $d_L$. One possible way to measure the inclination is through X-ray or radio observations of the associated gamma ray bursts (GRBs)~\citep{Guidorzi:2017ogy,Hotokezaka:2019}, but as noted above GRBs are only expected to be present in a fraction of the BNS mergers. Another possibility is though broadband photometry of the kilonova observations by analyzing the composition of the matter ejected. \cite{Dhawan:2019phb} successfully followed this route, and although their results were less precise by a factor of $\sim 4$ compared to the GRB measurements of \cite{Hotokezaka:2019}, it should be possible to replicate in a larger fraction of mergers. Therefore, even though it is still unclear which fraction of BNS could have EM inclination constraints and to which precision and accuracy, we will also consider a scenario where this information is incorporated via a prior on the inclination. In this scenario, we will consider a Gaussian prior centred at the true value and with a standard deviation of $10^{\circ}$, which matches the one found by~\cite{Dhawan:2019phb}.

Current and forthcoming wide-field telescopes, such as the Rubin Observatory (Rubin)\footnote{\url{https://www.lsst.org/}}, Wide Field Survey Telescope (WFST)\footnote{\url{https://wfst.ustc.edu.cn/}}, Multi-channel Photometric Survey Telescope (Mephisto)\footnote{\url{http://www.mephisto.ynu.edu.cn/site/}} and the Zwicky Transient Factory (ZTF)~\citep{Masci2019}\footnote{\url{https://www.ztf.caltech.edu/}}, will be able to detect a considerable fraction of kilonovae counterparts to BNS events which are well localised by the ET and/or CE. Since these GW detections are still more than a decade away, other telescopes and cameras may still come online in the meantime.

In our analysis we will start by comparing the capabilities of all four telescopes: Rubin, WFST, Mephisto and ZTF. Using the published sensitivities of each one, we compute the required exposure time making use of the power-law fits derived in~\cite{Zhu:2021ram} in order to use a fixed amount of the total available telescope time. This calculation follows the one performed for the case of Rubin by~\cite{Chen:2020zoq}. Fixing \emph{a priori} a fraction of telescope time dedicated to BNS GW follow-up allows a better comparison of all four facilities. Moreover,  Rubin scans mostly the southern hemisphere whereas the other telescopes cover the northern region, so their data will complement Rubin's. We quote results for two different fraction of telescope time ($F_{\rm time}$) needed to observe all the events. To wit, we assume either $F_{\rm time} = 10\%$ or $50\%$, with the latter being our baseline case. Although dedicating up to half of all usable telescope time for kilonova follow-ups may sound excessive, the third generation of GW detectors here considered are not expected to come online before the second half of the 2030s, therefore after the completion of the original proposed surveys for these instruments.

For each scenario, following~\cite{Chen:2020zoq} we select events with sky localization $\Delta\Omega\lesssim20\,\rm{deg}^{2}$ to ensure a small number of pointings needed to identify that event in the sky. We also make the conservative simplifying assumption that the average number of  pointings $p$ for these events will be given by
\begin{equation} \label{pointingseq}
    p = \frac{\rm 40 \; deg^2}{\rm FoV} \,,
\end{equation}
where FoV is the telescope effective field of view in squared degrees. The FoV values for each telescope used here are  shown in Table~\ref{tab:texp-comparison}. We also assume that the observable sky area for all four telescopes are equal in size, and that the fraction $f_{\rm obs}$ of events followed-up is given in all cases by $f_{\rm obs} = 0.4$. Although the observable sky area in each night is larger than this for all four telescopes, this lower number is supposed to already take into account unfavorable meteorological conditions and instrument downtime.  We remark that this is a conservative choice, and that other authors assume a more aggressive choice of $p = \Delta \Omega$/FoV~\citep{McGee:2018qwb}. Nevertheless this also takes into account the fact that the GW search area may be elongated which leads to wasted area in the borders of the FoV.

Finally, we consider two changes to the analysis of~\cite{Chen:2020zoq}: (i) we do not adopt a sharp cutoff in the detection horizon $d_{L,\rm{lim}}$, and instead consider the full extension of the efficiency curve without ignoring its smooth tail; (ii) for each efficiency curve we propose that the exposure time is adjusted so that less time is spent in closer events, which are brighter and easier to spot. One can thus maximize the number and distance of observable kilonovae while maintaining maximum completeness for the closer events. This advanced strategy will be better described in the next subsection.

\begin{figure}
	\centering
	\includegraphics[width=\columnwidth]{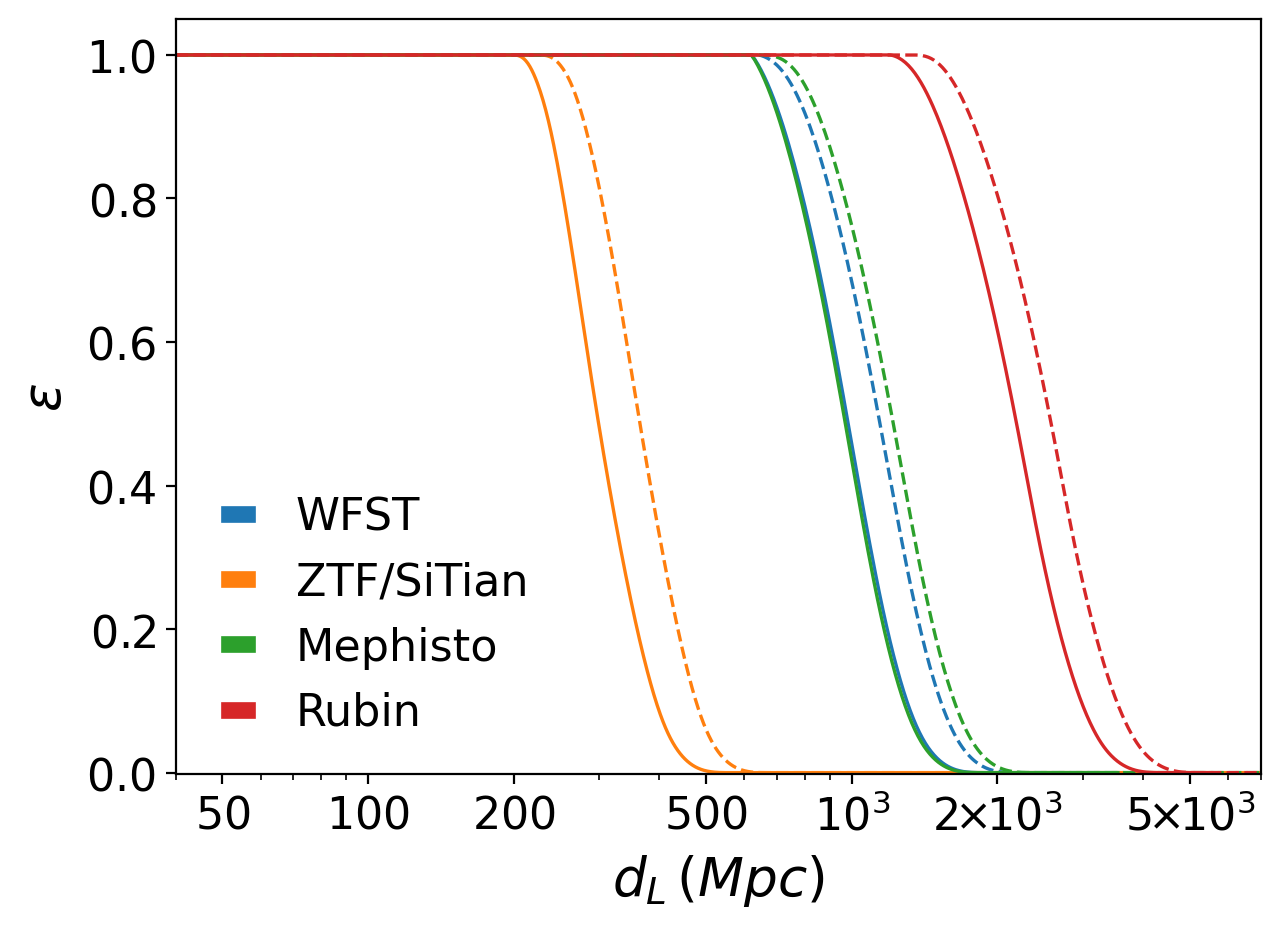}
	\caption{Kilonova efficiency $\varepsilon$ as a function of $d_{L}$ for the telescopes used in this work (Rubin, WFST, Mephisto and ZTF), and two exposure time: $t_{\rm exp.}=300\,\rm s$ (solid curve) and $t_{\rm exp.}=600\,\rm s$ (dashed curve).}
	\label{efficiencycurve}
\end{figure}

For the kilonovae program, we use the results presented in~\cite{Zhu:2021zmy}. They assume that all the simulated kilonovae are AT2017gfo-like, adopt a total ejecta mass of $M_{\rm ej}=0.04M_{\odot}$, a half-opening angle equal to $\Phi=60^{\circ}$, and ignore the viewing-angle dependence. In our program, we assume that the electromagnetic counterparts will be identified by three epochs of observations in two filters (\textit{gr}, that are typically used by the surveys we consider). Although other filters will be available that could be added to the analysis, we choose not to include them since the information added on the magnitudes PDFs would not compensate the increase of the observation time for the identification of the kilonova. The KN efficiency curves were found using the PDFs for the apparent magnitude and the absolute magnitude presented in ~\cite{Zhu:2021zmy}. For each band we convolve these PDFs to find the PDF $P_\mu$ of distance modulus $\mu$, and then find the luminosity distance PDF $P_{d_L}$:
\begin{align}
    P_{\mu} &\equiv P_{m_{\rm app}}\otimes P_{\left(-m_{\rm abs}\right)} \,,\\
    P_{d_L} &= \frac{5}{\left(d_L/{\rm pc}\right)\ln 10} P_{\mu} \,.
\end{align}
The kilonova detectability will basically depend on the filter selection and the survey sensitivity that is characterized through the limiting magnitude. All the limiting magnitude values were calculated using their relation to the exposure time presented in Table 2 of~\cite{Zhu:2021ram}. The KN efficiency curves are shown in Figure~\ref{efficiencycurve}. %The fraction of observed events in each scenario are presented in Table \ref{tableEM}.

% trim={<left> <lower> <right> <upper>}
\begin{figure*}
	\centering
	\includegraphics[trim={.2cm 0cm .2cm 0cm}, clip, width=2.05\columnwidth]{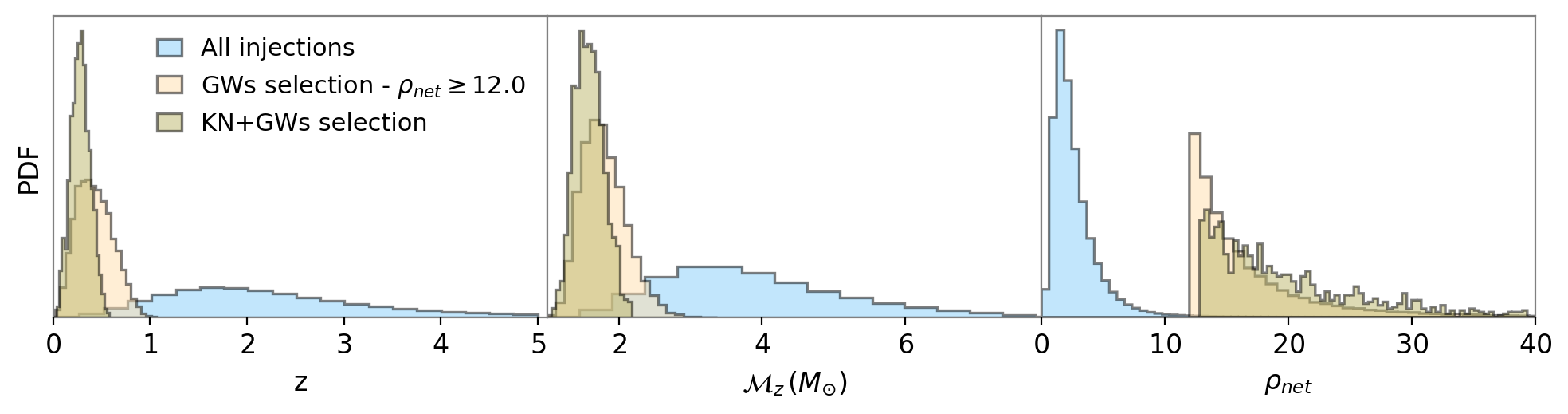}
	\caption{Distribution of a subset of GW parameters from our original catalog (blue curves) and after the GWs selection criteria adopted ($\rho_{\rm net}\geq12$ - beige curves) and EM counterpart selection considering $F_{\rm time}$ = 0.5 for the Rubin survey (see Table \ref{tab:Ftime} - olive green curves). Here $\mathcal{M}_z$ is the redshifted chirp mass defined as $\mathcal{M}_{z} = \left(1+z\right)\mathcal{M}$.}
	\label{SNRdistribution}
\end{figure*}

Our simulations predict 5498 BNS merger detections (events with $\rho_{\rm net}\geq 12$) per year with ET. This number represents only $\sim 1\%$ of our whole mock catalog ($z \le 5$). After imposing the cutoff for detection of an EM counterpart, this number is reduced according to the kilonova observation scenario (as will be shown below in Tables \ref{tab:texp-comparison} and \ref{tab:Ftime}). As an example, in Figure~\ref{SNRdistribution} we show the distribution of event parameters before and after imposing GW and EM selection cuts in a given scenario. The GW selection curves show that the detections occur in a reduced redshift range, as expected, selecting sources up to $z_{\rm lim}\sim 1.06$; this value shows that the choice of $z_{\rm max}=5$ as the threshold of the BNS GW distribution is enough to ensure that all relevant events are considered. This selection effect also impact the distribution of $\mathcal{M}_{z}$, that prefers events with low redshifted chirp mass. The distribution of $\rho_{\rm net}$ for all injections has a mean equal to $\approx 2.7$, and we can see that most of the events are distributed in $\rho_{\rm net}<10$ which explain the computed low fraction of BNS GW which are detected.

\subsection{Exposure time as a function of luminosity distance}

BNS coalescences that are closer to us require shorter exposure times to have their electromagnetic counterpart detected, when compared to similar events at larger distances. This implies that the dedicated exposure time should be a function that grows with the luminosity distance. Often, however, a simplified analysis disregards this fact by considering that the exposure time is constant for all kilonovae, regardless of their distances~\citep[e.g.][]{Chen:2020zoq}. Here we will propose instead that since BNS GW events themselves already constrain the luminosity distance, dedicated follow-up programs could use this information to adjust exposure times to be a function of the luminosity distance.

A simple estimation indicates that to maintain a constant signal-to-noise ratio, the exposure time should be proportional to $d_L^4$. The reason is that the flux (i.e. the signal) decreases with the square of the distance, and that the instrumental noise decreases roughly with the inverse square root of the exposure time. In practice, instrumental noise may not follow exactly this simple rule, but in any case the limiting magnitude of a given telescope as a function of exposure time is well understood. For the four telescopes here considered, \cite{Zhu:2021zmy} approximates the limiting magnitude in different bands as a power law of the exposure time. Making use of this we adjusted iteratively the exposure time in order to stretch the efficiency curves of the instruments to the maximum possible distance  $d_L^{\rm max}$, defined as the luminosity distance where the detection efficiency is larger than 99$\%$. This  guarantees that basically all kilonovae up to that distance will be detected.

\begin{figure}
	\centering
	\includegraphics[width=\columnwidth]{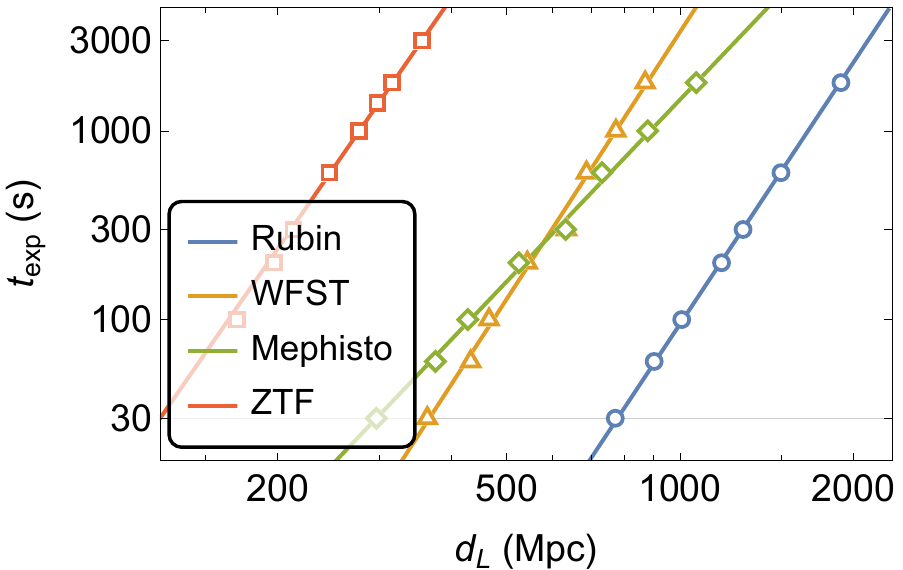}
	\caption{Required telescope exposure time $t_{\rm exp}$ as a function of the minimum luminosity distance with 99\% telescope efficiency for the surveys here considered. The data are well fitted by power-laws with exponents not too different from the naive expectation of 4 [see text]. We assume a minimum $t_{\rm exp}$ of 30s.}
	\label{fig:texp}
\end{figure}

We then fit a power law in the form:
\begin{equation} \label{eq:texp}
    t_{\rm exp} \,=\, a \, \left(\frac{\dd _L}{400 \, {\rm Mpc}}\right)^{n} \,,
\end{equation}
where the constants $\{a, n\}$ provide the best fit to the data. Figure~\ref{fig:texp} illustrates the result, showing that in all cases a power law is a very good fit.
Clearly a power-law exposure time program is problematic when the distances involved are too low or too high. When $d_L$ is too low, the total telescope time becomes dominated by the slewing time when changing pointings. For very far away sources, on the other hand, the time to be spent on a single object becomes prohibitive, not to mention the inherent challenges of performing very long exposures in astronomy. We therefore set a minimum and maximum exposure times: the former is fixed to be 30s, a standard exposure time in surveys, while the latter is set by the allocated fraction of yearly telescope time. We then have
\begin{equation} \label{eq:texp-final}
	t_{\rm exp}\left(d_L, t_{\rm exp}^{\rm max}\right) = \left \{
	\begin{array}{ccc}
	30\,s, & a d_L^{n} < 30\,s \\
	ad_L^{n}, & 30\,s<ad_L^{n}<t_{\rm exp}^{\rm max} \\
	t_{\rm exp}^{\rm max}, & ad_L^{n}>t_{\rm exp}^{\rm max}\\
	\end{array}
	\right.
\end{equation}
where $t_{\rm exp}^{\rm max}$ is the maximum exposure time.

\begin{table}
	\centering
	% \footnotesize
	%\setlength{\tabcolsep}{4.0pt}
	\begin{tabular}{c c c}
		\hline
		Telescope & $a (s)$ & $n$\\
		\hline
		Rubin     & 1.4  & 4.6 \\
		WFST      & 45   & 4.7 \\
		Mephisto  & 78   & 3.2 \\
		ZTF       & 5000 & 4.4 \\
		\hline
	\end{tabular}	
    \caption{The optimized exposure time parameters of Eq.~\eqref{eq:texp} for the different telescopes considered in this work.}
    \label{tablefit}
\end{table}

Table~\ref{tablefit} presents the values of $\{a,n\}$ that represent the best fit for each of the surveys. The exponents $n$ are not too far from the value 4, the simple expectation discussed above which assumes the limiting magnitudes scale as $1/\sqrt{t_{\rm exp}}$.

As expected, we find that adjusting the follow-up exposure time using the distance estimate from GW observations allows the telescope to spend less time in finding nearby (and brighter) events and longer times in those farther away. The total observing time $t_{\rm source}$ required per source was computed using:
\begin{equation}
    t_{\rm source} = p\, t_{\rm exp} N_{\rm epochs} N_{\rm filters} + 120s\,,
\end{equation}
where $t_{\rm exp}$ is the exposure time, $N_{\rm epochs}$ is the number of epochs, which we assume to be 3, $N_{\rm filters}$ is the required number of filters, which we assume to be 2 and \textit{p} is calculated by equation \eqref{pointingseq}. The last term is a two minutes overhead required for slewing and for filter changes, following \cite{Chen:2020zoq}. Table~\ref{tab:texp-comparison} compares a scenario in which fixed 300s exposure times are used with our proposal, fixing the maximum exposure in each case to ensure the observation of the same amount of kilonovae.  We assume that all surveys will have a total of 3600 hours available per year. In all cases, the same number of events are observed while consuming in between 20 and 30\% less telescope time. In what follows we will therefore assume the observational program described in Eq.~\eqref{eq:texp-final}.

\begin{table}
    \centering
    % \footnotesize
    \setlength{\tabcolsep}{3.5pt}
    \begin{tabular}{c c c c c c c}
    \hline
    Telescope & \! FoV \! & $f_{\rm obs}$  &  scenario & $N_{SS}/$yr & $F_{\rm time}$ \\
    \hline
    \multirow{2}{*}{Rubin}  & \multirow{2}{*}{9.6} & \multirow{2}{*}{0.4} &  $t_{\rm exp} = 300$s & \multirow{2}{*}{1194} & 0.66  \\
         &   &    & $t_{\rm exp} \propto d_L^{4.5}$ &  & 0.47  \\
    \multirow{2}{*}{WFST} & \multirow{2}{*}{6.6} & \multirow{2}{*}{0.4} & $t_{\rm exp} = 300$s  & \multirow{2}{*}{307}  & 0.26  \\
         &  &  & $t_{\rm exp} \propto d_L^{4.7}$ &   & 0.20  \\
    \multirow{2}{*}{Mephisto} & \multirow{2}{*}{3.1} & \multirow{2}{*}{0.4} &  $t_{\rm exp} = 300$s  & \multirow{2}{*}{298}  & 0.54  \\
     &  &  & $t_{\rm exp} \propto d_L^{3.2}$ &   & 0.46  \\
    \multirow{2}{*}{ZTF}  & \multirow{2}{*}{47}  & \multirow{2}{*}{0.4}  & $t_{\rm exp} = 300$s  & \multirow{2}{*}{11}   & 0.0015 \\
          &  &  & $t_{\rm exp} \propto d_L^{4.4}$  &    & 0.0012  \\
    \hline
    \end{tabular}
    \caption{\label{tab:texp-comparison}
    Comparison of different wide field telescope properties and of different exposure time strategies. FoV is the field of view in squared degrees, $f_{\rm obs}$ is the assumed fraction of the sky that observed, $N_{SS}/$yr is the number of gravitational waves that will have their counterparts observed per year, and $F_{\rm time}$ is the fraction of telescope time needed to observe all the events. As can be seen, adjusting the exposure time to be a power of the estimated distance decreases the needed amount of survey time.
    }
    \end{table}

With the observational method established, the next interesting forecast is on the total number of expected observed events for a fixed fraction of the total telescope time available each year. We compare two cases: using 10\% and 50\% of each telescope time. Table~\ref{tab:Ftime} shows the results. Clearly Rubin is the performance winner. Nevertheless, the other telescopes combined are still able to observe a good amount of events, albeit at lower redshifts. Since they cover the northern hemisphere it shows that they may perform an interesting complementary follow-up to Rubin in this hemisphere.

Dedicating a large fraction of telescope time to observe transients such as kilonovae with long expositions may lead to some events being lost due to their overlap in the time domain. In Appendix~\ref{app:overlap} we investigate this and show that with the possible exception of ZTF, in all other cases this would be a negligible effect.

\section{Cosmological Forecasts} \label{sec:forecasts}

In this section we will discuss the advantages of using KN measurements as tracers of both density and velocity fields, besides only as distance indicators. Since EM counterparts from BNS are expected mostly in $z\le0.5$, as can be seen in Figure~\ref{SNRdistribution}, we will split our forecasts in lower redshift ($z\le0.5$) and intermediate/high redshifts ($z\ge0.5$). This is a simple yet useful separation in light of the recent framing of the Hubble tension as a tension between lower and higher redshifts~\citep[see e.g. the review][]{Abdalla:2022yfr}. We will assume as our baseline survey a 5-year observational run of the ET and the follow-up facilities.

\begin{table}
\centering
\setlength{\tabcolsep}{2.0pt}
\begin{tabular}{c c c c c c c}
	\hline
	Telescope & $t_{\rm exp}^{\rm max}(\rm s)$  & $z_{\rm max}$ & $f_{20\,\rm{deg}^{2}}$ & $f_{\rm obs}$ & $N_{SS}$/yr & $F_{\rm time}$ \\
	\hline
	Rubin    & 90 & 0.49 & 0.89 & 0.4 & 819 &  0.1  \\
	WFST     & 200 & 0.27 & 0.94 & 0.4 & 244 &  0.1 \\
	ZTF      & 3200 & 0.17 & 0.98 & 0.4 & 78 & 0.1  \\
	Mephisto & 140 & 0.23 & 0.96 & 0.4 & 157 &  0.1 \\
	\hline 	
	Rubin & 315 & 0.61 & 0.86 & 0.4 & 1196 &  0.5 \\
	WFST    & 550 & 0.34 & 0.92 & 0.4 & 404 &  0.5  \\
	ZTF     & 8500 & 0.22 & 0.96 & 0.4 & 152 &  0.5 \\
	Mephisto & 320 & 0.30 & 0.93 & 0.4 & 306 &  0.5 \\
\hline
\end{tabular}
\caption{Similar to Table~\ref{tab:texp-comparison} but for the same fixed fractions of survey time. Here $f_{20\rm deg^{2}}$ is the fraction of GW sources that have localization area < 20 $\rm deg^2$, and $t_{\rm exp}^{\rm max}$ is the maximum exposure time.
\label{tab:Ftime}
}
\end{table}

We use the symbols $g$ for galaxies and $c$ for standard candles in general. When we need to specify which standard candle we will use $s$ for supernovae and $k$ for kilonovae (assumed to be measured jointly with the associated GW). From standard sirens we extract directly the  luminosity distance $d_L(z_t)$. The observed redshift is the sum of a cosmological redshift $\bar{z}$ plus a small redshift induced by peculiar velocity: $1+z_{t}=(1+\bar{z})(1+z_p)\to z_{t}=\bar{z}+(1+\bar{z})z_p$, with $v_c=z_p$.  Then we can write
\begin{equation}
    d_L(z_{\rm obs}) \approx d_L(\bar{z})+\frac{\dd d_L}{\dd z}(1+\bar{z})v_c  \,,
\end{equation}
where $\bar{z}$ is the cosmological redshift. Therefore
\begin{equation}
    v_c=\frac{\Delta d_L}{d_L} \frac{1}{\dd\log d_L/\dd\log(1+z)}\,.
\end{equation}
As stated above, we will neglect the specificities of the beaming effect for GWs discussed in \citep{Torres-Orjuela:2018ejx,Torres-Orjuela:2020dhw} and model the velocity effects like that on photons. In the CMB rest frame (i.e. neglecting our own peculiar velocity) we have that~\citep{hui2006,Davis:2010jq}
\begin{equation}
    d_L(z_{\rm obs}) \approx d_L(\bar{z})(1-2v_c)+\frac{\dd d_L}{\dd z}(1+\bar{z})v_c \,,
\end{equation}
which leads to
\begin{equation}
    v_{\rm c}=\frac{\Delta d_L}{d_L}\left[\frac{\dd\log d_{L}}{\dd\log(1+z)} - 2\right]^{-1}\label{eq:magn-vel} \,.
\end{equation}
The statistical uncertainty in the velocity field is thus~\citep{Amendola:2019lvy}:
\begin{equation}
    \sigma_{v,{\rm eff}}^{2}\!\equiv\!\left[\frac{\Delta d_L}{d_L}\right]^{2}\!\left[2-\frac{\dd\log d_{L}}{\dd\log(1+z)}\right]^{-2}\!\!\!+\frac{\sigma_{v{\rm ,nonlin}}^{2}}{c^{2}}.\label{eq:new-sv-1}
\end{equation}
Since we assume for our fiducial case $\Omega_{k0}=0$, we can write $d\log d_{L}/d\log(1+z) = 1 + (1+z)^2/[d_L H(z)]$. We also take $\sigma_{v{\rm ,nonlin}} = 300$ km/s, although this parameter has little impact in practice.

We have therefore three random fields: the already mentioned (radial) velocity of the standard candles $v_c$, and the density fields of both galaxies ($\delta_g$) and of the standard candle ($\delta_c$). We use subscripts $v,g,c$ to refer to these three fields, respectively. These give rise to six linear auto- and cross-power spectra. We define as $b_i$ the linear bias of a given tracer $i$ of the dark matter density-contrast field $\delta$ and the linear growth rate $f$ as
\begin{equation}
    f \equiv \frac{\dd\log\delta}{\dd\log a} = -\frac{\dd\log D_+(z)}{\dd\log (1+z)} \simeq \Omega_{m}(z)^\gamma \,,
\end{equation}
where  $D_+$ is the growth function and $\gamma$ is the growth-rate index, which is assumed constant as a simple parametrization for the linear growth which allows beyond General Relativity behaviour (the GR case corresponds to $\gamma = -0.545$). We then have~\citep{Quartin:2021dmr}
\begin{align}
     \!P_{\rm gg}(k,\mu,z) \!&= \Upsilon\big[1+ \beta_{\rm g} \mu^{2}\big]^2 \,b_{\rm g}^{2} \,S_{\rm g}^2\, D_+^2 P_{\textrm{mm}}(k) + \frac{1}{n_{\rm g}}, \label{eq:pgg} \\
     \!P_{\rm cc}(k,\mu,z) \!&= \Upsilon\big[1+ \beta_{\rm c} \mu^{2}\big]^2 \,b_{\rm c}^{2}\,S_{\rm c}^2 \, D_+^2 P_{\textrm{mm}}(k) + \frac{1}{n_{\rm c}}, \label{eq:pss} \\
     \!P_{\rm gc}(k,\mu,z) \!&= \Upsilon\big[1+ \beta_{\rm g} \mu^{2}\big]\big[1+ \beta_{\rm c} \mu^{2}\big] \,b_{\rm g} \,b_{\rm c}\,S_{\rm g} \,S_{\rm c} \, D_+^2 P_{\textrm{mm}}(k) \nonumber\\
     & \quad\; + \frac{n_{\rm gs}}{n_{\rm g}n_{\rm c}}, \label{eq:pgs} \\
     \!P_{\rm gv}(k,\mu,z) \!&= \!\Upsilon\frac{H\mu}{k(1+z)} \!\big[1 + \beta_{\rm g}\mu^{2}\big] b_{\rm g} S_{\rm g} S_{\rm v}  f  D_+^2 P_{\textrm{mm}}(k), \label{eq:pgv} \\
     \!P_{\rm cv}(k,\mu,z) \!&=\!\Upsilon\frac{H\mu}{k(1+z)} \!\big[1 + \beta_{\rm c}\mu^{2}\big] b_{\rm c} S_{\rm c} S_{\rm v} \, f  D_+^2 P_{\textrm{mm}}(k), \label{eq:psv} \\
     \!P_{\rm vv}(k,\mu, z) \!&= \Upsilon\!\left[\frac{H\mu}{k(1+z)}\right]^2 S_{\rm v}^2 \,f^{2} \,D_+^2 P_{\textrm{mm}}(k)+ \frac{\sigma^2_{v, {\rm eff}}}{n_{\rm c}},
    \label{eq:pvv}
\end{align}
where $\Upsilon \equiv (H d_{L,r}^2)/(H_{r}d_{\rm L}^2)$, $\beta_i \equiv f/b_i$, $\mu \equiv \hat{k} \cdot \hat{r}$, with $i=g,c$,  $S_{\rm g,c,v}$ are damping terms, $n_{g,c}$ are the number densities of galaxies and sirens, $n_{gc}$ is the fraction of sirens in galaxies belonging to the survey (here assumed to vanish for simplicity), and $P_{\textrm{mm}}$ is the matter power spectrum at $z=0$. We also take into account the fact that the values of $k$ and $\mu$ depend on the cosmological model (the Alcock-Paczynski effect) through $H$ and $d_L$ \citep[for more details, see][]{Amendola:2019lvy}. As in \cite{Quartin:2021dmr},  the smoothing factors are modeled as
\begin{equation}
    S_{\rm g,c,v}=\exp\left[-\frac{1}{4}(k\mu\sigma_{\rm g,c,v})^{2}\right],
\end{equation}
with fiducial values chosen as $\sigma_{\rm g}=\sigma_{\rm c} = 4.24 \;{\rm Mpc}/h$ and $\sigma_{\rm v}= 8.5$ Mpc$/h$ (see \cite{2014MNRAS.445.4267K,Howlett:2017asw,Dam:2021fff} for the choice of these values).

The full analysis using all six spectra for supernovae was dubbed the \six\ $g$--$s$--$s$ method in~\cite{Quartin:2021dmr}. For sirens we can instead refer to it as a \six\ $g$--$k$--$k$ method ($k$ as a shorthand for KN or any other standard siren), but we will often  refer to it as simply \six\ for short. Likewise, using sirens only as velocity but not density tracers (i.e., dropping the $P_{\rm cc}$, $P_{\rm gc}$ and $P_{\rm cv}$ terms), we have a \three\ $g$--$k$ method (or simply \three\ for short). Finally, using only $P_{\rm gg}$ consists of a traditional full shape galaxy power spectrum approach. We remark that if one relies on GW without electromagnetic counterparts, another possibility is to use the GW in luminosity-distance space, as opposed to redshift space. This entails a joint analysis of clustering similar to the one proposed here, but in this case the GW part exhibits no RSD terms. This was in fact proposed by~\cite{Mukherjee:2020hyn} and \cite{Diaz:2021pem}, where a
\three\ $g$--$gw$ analysis was carried out and constraints on background cosmological parameters were forecast.

For the sirens, the distance error estimates were obtained with the method discussed in Section~\ref{sec:gw-params}, both including or not EM information on the BNS orbit inclination (through the addition of a prior).  For supernovae, which we will also employ to forecast performance as distance estimators for comparison, we assume distances are measured with magnitude uncertainties given by the sum in quadrature of the intrinsic scatter $\sigma_{\rm int} = 0.13$ mag with the lensing-induced scatter of $\sigma_{\rm lens} = 0.052 z$~\citep{Quartin:2013moa}. For the number of events for lower redshifts we follow~\cite{Quartin:2021dmr} assuming 15\% SN completeness, for a total of 239k SN. For higher redshifts we use the expected number of SN with the Nancy Grace Roman Space Telescope as computed by~\cite{Rose:2021nzt}. Finally, for the CMB we quote the results found in~\cite{Quartin:2021dmr}, which comes from a combination of Planck 2018 TTTEEE~\citep{Planck:2018vyg} without lensing and an analysis on $\sigma_8$ and $\gamma$ performed by~\cite{Mantz:2014paa}.

For galaxies, we produce two different forecasts. For $z\le0.5$ we  use the predicted values for the DESI Bright Galaxy Survey presented in \cite{DESI:2016fyo}, which has an expected linear bias of $b_{g} = 1.34/D_+(z)$. Converting their numbers to volumetric density, we find $n_{g} = $ \{38.50, 17.63, 6.439, 1.937, 0.3571\} $10^{-3}(h/{\rm Mpc})^3$ in the \textit{z} bins of $\Delta z=0.1$ centered on \{0.05, 0.15, 0.25, 0.35, 0.45\}, respectively. For $0.5<z<1.5$ we will forecast the performance of the traditional \one\ using only galaxies. We will use as baseline the DESI Emission Line Galaxy (ELG) survey, which covers well this redshift range. The assumed linear bias in this case is $b_{g} = 0.84/D_+(z)$~\citep{DESI:2016fyo}, and the number densities are \{0.1778, 1.099, 0.8130, 0.7940, 0.5007, 0.4384, 0.4097, 0.1532, 0.1316\} $10^{-3}(h/{\rm Mpc})^3$ for the bins centered on \{0.65, 0.75, 0.85, 0.95, 1.05, 1.15, 1.25, 1.35, 1.45\}.  Finally, since the kilonova bias as density tracers has not been studied in detail, we assume for simplicity that $b_{\rm k} = 1.0/D_+(z)$, and likewise for SN.

The \six\ Fisher matrix (FM) for a set of parameters $\theta_{\alpha}$, in a survey of volume $V$ and for an interval $\Delta_k$ of $k$-modes, is~\citep{Tegmark:1997rp,Abramo:2019ejj}
\begin{equation}
    F_{\alpha\beta} \,=\, \frac{1}{(2 \pi)^3} 2\pi k^{2}\Delta_{k}V\bar{F}_{\alpha\beta} \,=\, VV_{k}\bar{F}_{\alpha\beta}\,,
\end{equation}
where $V_{k}= (2\pi)^{-3}2\pi k^{2}\Delta_{k}$ is the volume of the Fourier space integrated over the azimuthal angle but not over the polar angle, and where
\begin{equation}
    \bar{F}_{\alpha\beta} = \frac{1}{2}\int_{-1}^{+1} \dd\mu \, \frac{\partial C_{ab}}{\partial\theta_{\alpha}}C_{ad}^{-1}\frac{\partial C_{cd}}{\partial\theta_{\beta}}C_{bc}^{-1} \,,
\end{equation}
to be evaluated at the fiducial value. The elements of the data covariance matrix $C_{ab}$, with $a,b$ standing for $g,c,v$, are the six power spectra \eqref{eq:pgg}--\eqref{eq:pvv}~\citep{Quartin:2021dmr}. Denoting with $V(z)$ the volume of the $z$-shell, the $k$-cells are chosen with size $\Delta_{k}=2\pi/V(z)^{1/3}$ between $k_{\rm min}(z)$ and $k_{\rm max}$. Following~\cite{Garcia:2019ita}, we take $k_{\rm min}=2\pi/V(z)^{1/3}$.

The choice of $k_{\rm max}$ is more delicate.  The amount of information grows rapidly with $k_{\rm max}$ but so do the modeling uncertainties due to non-linear effects. Ideally, one would like to select the highest $k_{\rm max}$ that does not introduce significant non-linear effects: this, however, depends clearly on the cosmological model. In \cite{Amendola:2022vte}, the dependence of Fisher matrix forecasts for a non-linear power spectrum on $k_{\rm max}$ has been explored. The conclusion was that, if one does not employ a specific cosmological model, one needs to have independent strong prior constraints on the linear and non-linear bias parameters to decide which $k_{\rm max}$ is safe to use. In this paper, however, we restrict ourselves to $\Lambda$CDM (but with a free growth index $\gamma$) and in this case the standard choice $k_{\rm max} = 0.1~h/$Mpc is likely to be sufficient to ensure we remain within the linear regime. This is also a value close to $0.08\,h/$Mpc, which was found in retrospect to be the one for which Fisher matrix precision calculations best matched that of current real data analysis~\citep{Foroozan:2021zzu}. Note also that since non-linearities are stronger at lower redshifts, one is expected to be able to reach higher $k_{\rm max}$ at higher redshifts~\citep{Nishimichi:2008ry,Tomlinson:2022xud}, but here we adopt a fixed value for simplicity.

The fiducial values of the parameters that are varied in the \six\ Fisher matrix  are
\begin{equation}
    \{\sigma_8,\,\gamma,\, H_0,\, \Omega_{m0},\, \Omega_{k0}\} = \{0.83,0.545, 70\,{\rm km/s/Mpc}, 0.3, 0\} .
\end{equation}
We also fix $n_s=0.96$ and $\tau=0.066$. The matter power spectrum is computed using CAMB\footnote{\url{https://camb.info/}}~\citep{Lewis:1999bs}. When analysing the combined constraints of clustering and distance methods, we simply sum the \six\ FM with the usual distance FM for KN or SN~\citep{2010deto.book.....A}. For KN, the distance parameters are just the background ones, to wit $\Omega_{m0},\, \Omega_{k0},\, H_0$, while for SN we have an extra parameter $M_B$ to account for their unknown absolute magnitude. We marginalize over this when combining with the clustering FM, but for the distance constraints alone this makes the SN FM degenerate. So, for this case alone, instead of $M_B$ and $H_0$ separately we use one single parameter which accounts for the combination $M_B - 5 \log H_0$ and marginalize over it. In practice, this can be achieved by simply fixing $M_B$ and marginalizing over $H_0$ (or vice-versa).

\begin{table}
    \centering
    % \footnotesize
    \setlength{\tabcolsep}{2.2pt}
    \begin{tabular}{l c c c c c}
    \hline
    $1\sigma$ uncertainties in:  & $\sigma_8$ & $\gamma$ & $H_0$ & $\Omega_{m0}$ & $\Omega_{k0}$  \\
    \hline
    \hline
    \multicolumn{6}{c}{Low $z$ ($0\le z \le 0.5$)} \\
    \hline
    \hline
    DESI BGS $gg$ & 0.081  & 0.165 & 2.1  & 0.0095 & 0.171  \\
    \hline
    Rubin $3\times2$pt g--k        & 0.079  & 0.137 & 2.1  & 0.0094 & 0.168  \\
    Rubin \six\ g--k--k            & 0.070  & 0.129 & 2.1  & 0.0093 & 0.167  \\
    Rubin \six\ + $\iota$ prior    & 0.070  & 0.127 & 2.1  & 0.0093 & 0.166  \\
    Rubin full-sky (``FS'') \six                 & 0.057  & 0.105 & 1.6  & 0.0067 & 0.127  \\
    \hline
    Rubin BNS distances            & -      & -     & 0.12 & 0.24  & 0.41  \\
    Rubin BNS dist + $\iota$ prior & -      & -     & 0.097& 0.18  & 0.31  \\
    Rubin BNS dist + \six          & 0.069  & 0.128 & 0.085& 0.0063 & 0.018  \\
    Rubin \emph{as above} + $\iota$ prior & 0.069 & 0.126 & 0.071& 0.0062 & 0.015  \\
    Rubin FS BNS \!dist \!+\! $6\!\times\!2$pt          & 0.056  & 0.104 & 0.060& 0.0046 & 0.013 \\
    \hline
    \end{tabular}
    \caption{\label{tab:1Derrors}
    Marginalized absolute forecast uncertainties in each cosmological parameter for different instrument capabilities and 5 years of observation. The combination of clustering measurements using BNS to their distances shrinks $H_0$ uncertainties by around 30\%, and constrains curvature to within 2\%.
    Rubin full-sky assumes a second equivalent Rubin coverage in the north, for a total coverage of $80\%$ of the sky. All results assume $k_{\rm max} = 0.10 \,h^{-1}$Mpc. $H_0$ is given in units of km/s/Mpc.
    }
\end{table}

Following~\cite{Quartin:2021dmr} we also employ three global nuisance parameters to account for the non-linear RSD ($\sigma_{g},\,\sigma_{c},\,\sigma_{v}$) and two bias nuisance parameters in each redshift bin ($b_{g}^{z_{i}}$ and $b_{c}^{z_{i}}$). We adopt a Gaussian prior with 0.5 uncertainties for $\Omega_{m0}$, $\Omega_{k0}$, and with 50\% errors in all bias and RSD damping parameters. For $\sigma_{8},\, \gamma$ and $H_0$ we use uninformative priors.

\begin{table}
    \centering
    % \footnotesize
    \setlength{\tabcolsep}{2.2pt}
    \begin{tabular}{l c c c c c}
    \hline
    $1\sigma$ uncertainties in:  & $\sigma_8$ & $\gamma$ & $H_0$ & $\Omega_{m0}$ & $\Omega_{k0}$  \\
    \hline
    \hline
    \multicolumn{6}{c}{Low $z$ ($0\le z \le 0.5$)} \\
    \hline
    \hline
    %Rubin \six                           & 0.070  & 0.129 & 2.1  & 0.0093 & 0.167  \\
    Rubin BNS distances                   & -      & -     & 0.12 & 0.24   & 0.41  \\
    Rubin BNS dist\,+\,$6\!\times\!2$pt\,g--k--k  & 0.069  & 0.128 & 0.085& 0.0063 & 0.018  \\
    %\hline
    Rubin SN distances             & -    & -     & -      & 0.050  & 0.092 \\
    %Rubin SN\,dist\,+\,$6\!\times\!2$pt\,g--k--k   & 0.069  & 0.128 & 1.5  & 0.0072 & 0.014  \\
    Rubin SN\,dist\,+\,$6\!\times\!2$pt\,g--s--s   & 0.065  & 0.115 & 1.5  & 0.0068 & 0.013  \\
    \hline
    \hline
    \multicolumn{6}{c}{Intermediate and high $z$ ($0.5\!\le\! z\! \le\! 1.5$)} \\
    \hline
    \hline
    DESI ELG $gg$                  & 0.032  & 0.181 & 1.4  & 0.0057 & 0.066  \\
    Roman SN distances             & -      & -     & -    & 0.084  & 0.232  \\
    DESI ELG $gg$ \!+\! Roman SN   & 0.031  & 0.181 & 0.98 & 0.0048 & 0.022  \\
    \hline
    \hline
    \multicolumn{6}{c}{Last scattering surface} \\
    \hline
    \hline
    Planck CMB (${}^\star$)               & 0.18  & 0.34  & 3.7   & 0.064 & 0.017 \\
    \hline
    \end{tabular}
    \caption{\label{tab:BNS-comparisons}
    Similar to Table~\ref{tab:1Derrors}, but comparing the performance with BNS with SN and with probes in other redshift ranges. SN forecasts assume a spectroscopic catalog in the same area of the sky. The higher number densities make supernovae better probes of relative distance and clustering, but KN are capable of measuring $H_0$ much better.
    }
\end{table}

Table~\ref{tab:1Derrors} summarizes our forecasts for 5 years of observation. All uncertainties are fully marginalized over all other parameters. Results are divided into three redshift ranges. For lower redshifts, where sirens will be useful, we compare the traditional full-shape (linear) power spectrum analysis with those of clustering combining galaxies and sirens, and with standard candle distance estimates. Additionally, we also separate the results to include (i) the contribution of the $\iota$ prior on the measurement of $d_{L}$, and (ii) assuming the scenario in which a second Rubin-like survey operate simultaneously in the northern hemisphere, doubling the observed sky area and consequently the number of galaxies and standard sirens measured. For the former, we find that the inclusion of a $10^\circ$ $\iota$ prior on the \six\ has only a very small effect. In fact, the uncertainty on $\gamma$ decreases by only $1.5\%$, even though such a prior results in improvements on their distance measurements by over $15\%$. In Appendix~\ref{app:sint} we explore why this is the case. Nevertheless for their use as traditional distance indicators, such a prior does lead to important improvements, reducing uncertainties by 20--25\% in each of the three background parameters.

Due to the expected low volume density of these objects, compared to traditional galaxy (full shape) power spectrum measurements, the performance gains achieved when combining them with density and velocity spectra measurements from KN are only modest. Peculiar velocity from sirens reduce modestly uncertainties in $\sigma_{8}$ and $\gamma$ ($\simeq5\%$ and 16\%, respectively). The addition of sirens as density tracers further decreases the uncertainties in $\gamma$ and $\sigma_{8}$ by almost 10\%. Since sirens provide absolute distance indicators, they can also measure $H_0$ as is well known. The use of sirens simply as distance measurements results in better precision in $H_{0}$ than the \six\ method. Nevertheless, the combined measurements of spectra and distances is able to reduce the uncertainties in  $H_{0}$ by a significant 30\%. Moreover, such a combination improves the measurements on curvature by an order of magnitude, and reaches the level of 2\% in $\Omega_{k0}$.

Table~\ref{tab:BNS-comparisons}  compares the performance of sirens with other probes, in particular with those of type Ia supernovae at the same redshift range, with supernovae and clustering at higher redshifts, and with the CMB constraints. At intermediate/high redshifts as discussed before we rely on DESI ELG forecasts for clustering, and the Roman Telescope for supernovae. Finally, we also show the CMB constraints. As can be seen, when combining clustering and distances, KN can outperform SN in $H_0$ due to their capabilities of measuring absolute distances. For the other cosmological parameters, SN perform similarly, with their higher number densities compensating for the extra degeneracies with $H_0$. In any case it will be very useful to have two independent distance probes with comparable performance potential, as this will be of great use in highlighting any possible systematics.

\begin{figure*}
	\centering
	\includegraphics[width=1.60\columnwidth]{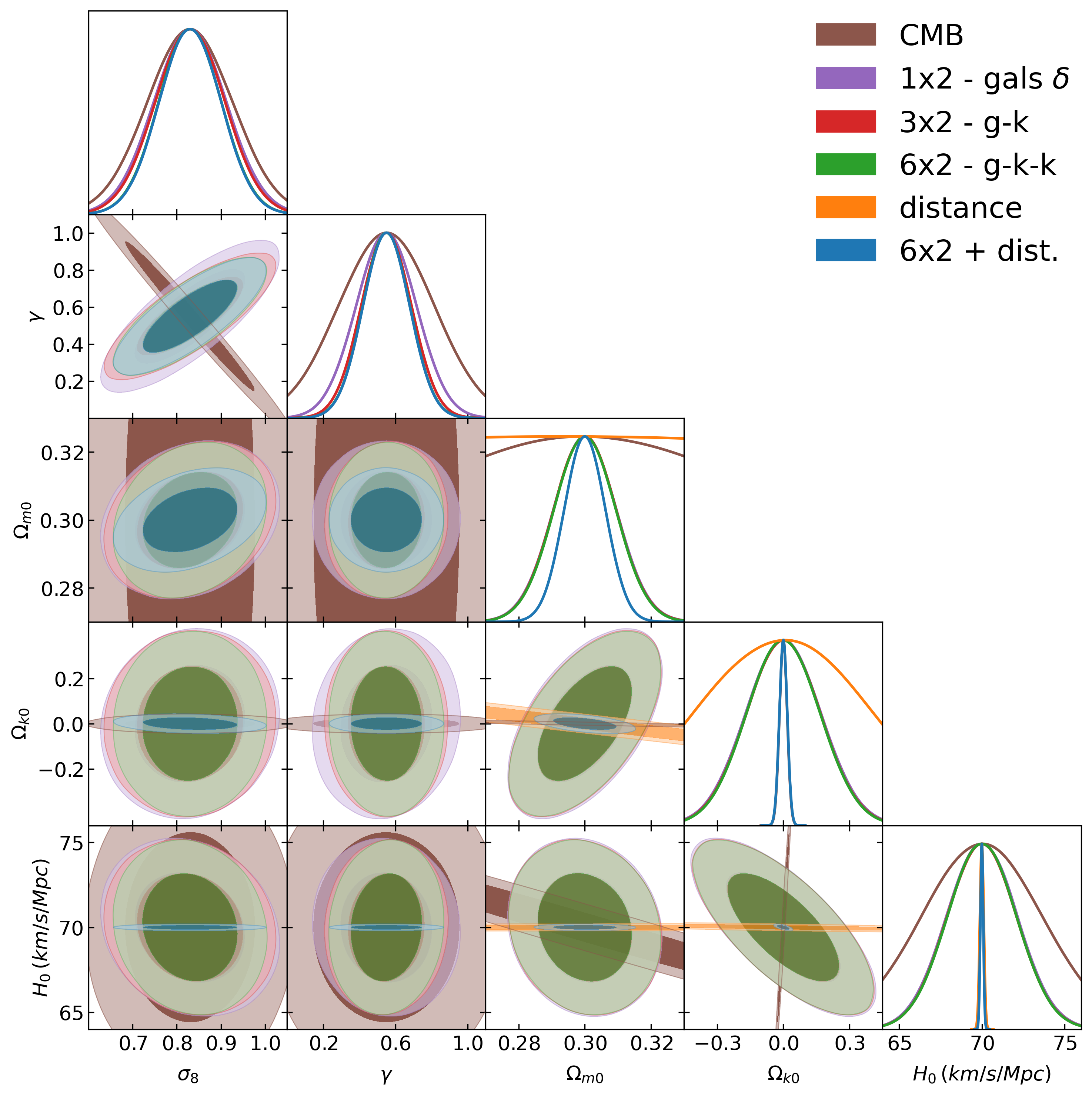}
    \caption{Marginalised contours for $\{\sigma_{8}, \gamma, \Omega_{m0}, \Omega_{m0}, H_{0}\}$, assuming $k_{\rm max}=0.1\, \rm{h/Mpc}$ and $z_{\rm max}=0.5$. Results are separated according to the method used. Purple: \one\ of galaxies; red: \three; green: \six; orange: siren distances only; blue: \six\ + distances; Brown: CMB.
	}
	\label{cornerplot}
\end{figure*}

\begin{figure}
	\centering
    \subfigure{\includegraphics[width=\columnwidth]{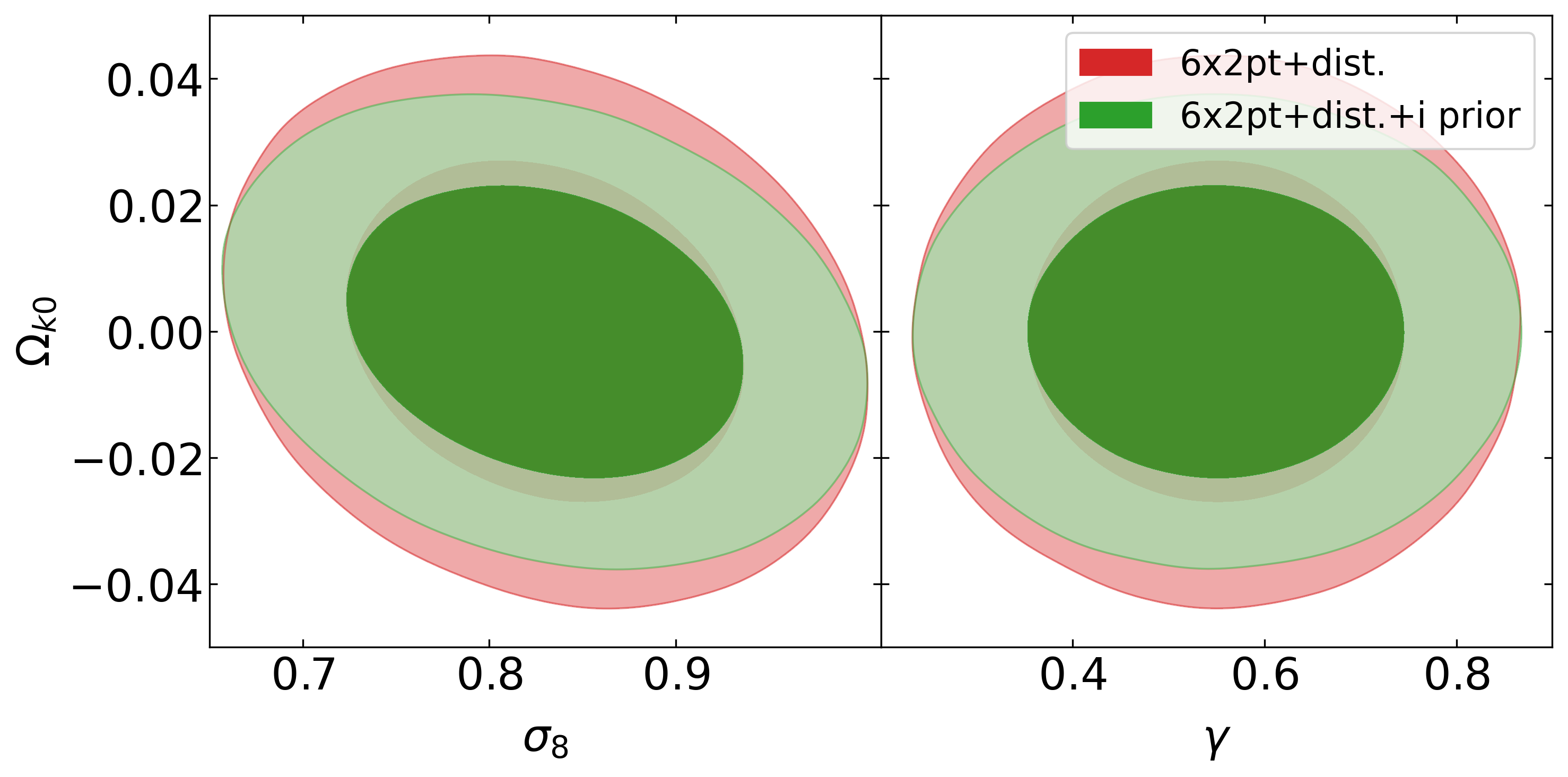}}
    \subfigure{\includegraphics[width=\columnwidth]{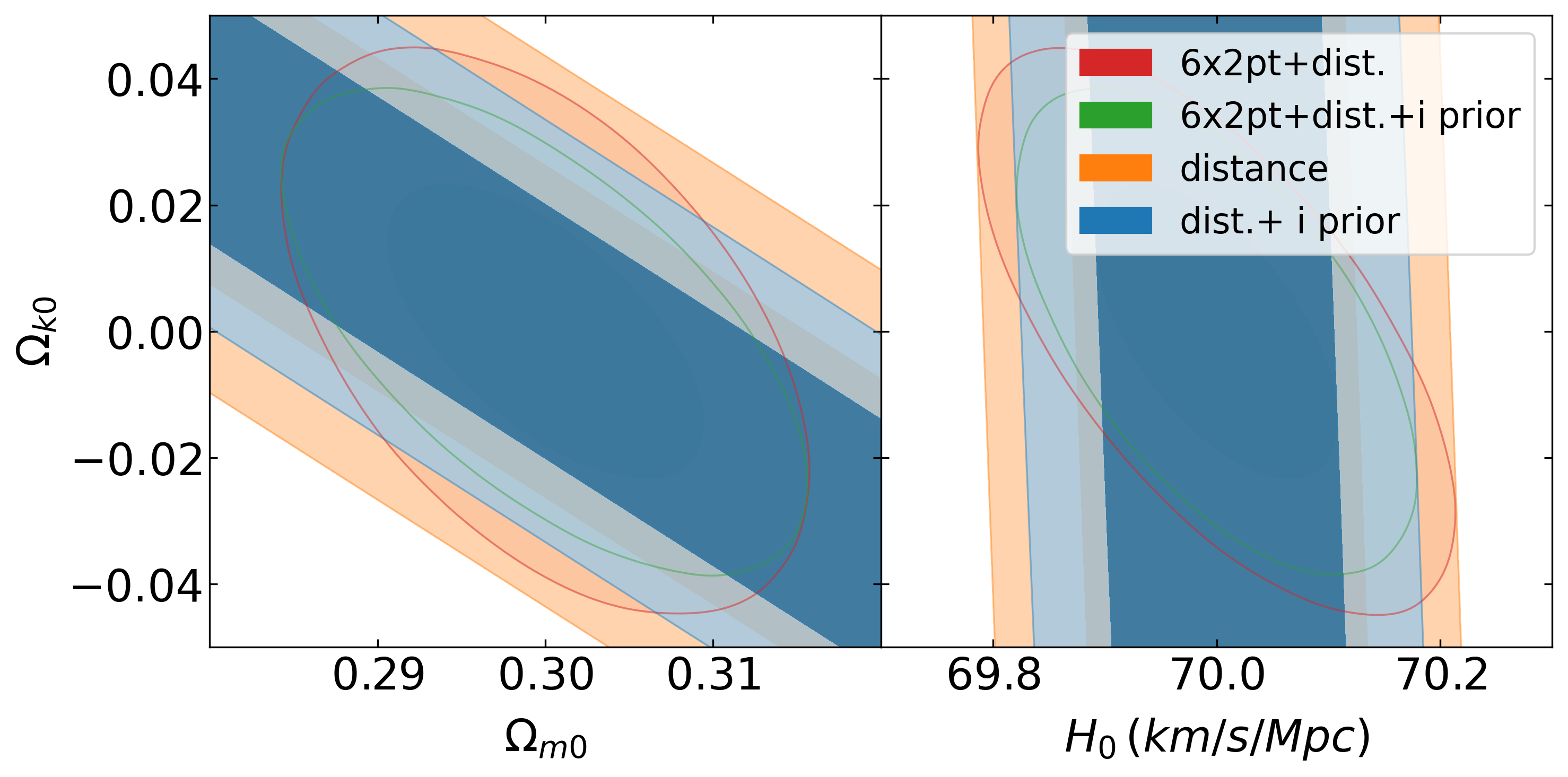}}
    \caption{Possible improvements due to the addition of external information on the BNS inclination.}
    \label{nsprioreffec}
\end{figure}

Figure \ref{cornerplot} illustrates the forecasts above by showing the 1$\sigma$ and 2$\sigma$ confidence regions for different parameter combinations. We depict different methods:  1$\times$2pt applied to galaxies as matter tracers, 3$\times$2pt, 6$\times$2pt, the use of the siren distances alone (the usual Hubble diagram fit), the combination of distances and \six\ and Planck TTTEEE CMB results.  We see that the influence of peculiar velocities is greater on $\gamma$ and $\sigma_{8}$, while in the other parameters ($\Omega_{m0}, \Omega_{\kappa0}, H_0$) the effect is minimal.
The same occurs when comparing the 3$\times$2pt and 6$\times$2pt methods, where we see a modest improvement over $\gamma$ and $\sigma_{8}$  (the 2D figure-of-merit in the $\sigma_{8}-\gamma$ plane increases by 12\%), while the other parameters have almost identical results.

It is also interesting to note that although siren distances alone produce highly degenerate constraints in the $\Omega_{m0}-\Omega_{k0}$ plane, this is distinctively different from the degeneracies on the clustering measurements, and this high complementarity result in very tight constraints in the background parameters when combining both methods. This in turn leads also to modest improvements in the perturbation forecasts from the \six\ baseline, and to the 30\% improvements in $H_0$ from the distance-based results mentioned above. We also remark the almost orthogonal degeneracies exhibited by both the distance method and the CMB, in the $H_0-\Omega_{k0}$ plane, and by the \six\ method and the CMB in the $\sigma_8-\gamma$ plane. The latter was first noticed by~\cite{Quartin:2021dmr} in the case of supernovae, but the former is particular to standard sirens.

Finally, Figure~\ref{nsprioreffec} illustrates the effect of including external EM information on the inclination $\iota$ (as a $10^\circ$ prior) on the cosmological results. The $\iota$ prior reflects this possible additional information from the electromagnetic counterpart, which helps to break the inclination angle–distance degeneracy, and consequently reduces the uncertainty of $d_{L}$. This induces improvements on the uncertainties of $H_{0}$, $\Omega_{m0}$, and $\Omega_{\kappa0}$ found with the distance method of 19\%, 25\%, and 24\%, respectively. The effect on the \six\ method alone is almost negligible, as discussed above, so this is not depicted. But still we get interesting improvements for the combined \six\ + distances case, as can be seen.

\section{Discussion} \label{sec:discussions}

The next generation of ground-based GW detectors predict to detect gravitational waves from BNS at cosmological distances. Here we revisited their use as observables both in the traditional distance measurements as well as in the use as tracers of the velocity and density field. We consider that there are three main results of our work.

First, we propose a method to optimize the use of telescope time in BNS GW follow-ups, which uses the preliminary GW information on the distance to the source to allocate telescope time accordingly. This means longer integration times on further objects, and can result in a reduction of needed telescope time by 20--30\% for the same number of kilonova observations in a given redshift range. This is relevant as by the next generation of GW facilities the number of observed standard sirens will be limited by telescope time.

Second, we find that using clustering in the full-shape power spectrum method, using sirens both as velocity and density tracers, their expected low number density results for the sirens means that the gains are smaller than those obtained when using supernovae as tracers. Nevertheless, we obtain modest gains on the growth of structure precision when compared to the constraints arising purely from galaxy clustering. To wit, we obtain a 14\% (24\%) reduction on the uncertainties in $\sigma_8$ and $\gamma$.

Finally, we show that the combination of distance and clustering of standard sirens (using the full \six\ method) results in massive improvements in the amount of background cosmological information that can be inferred. For the curvature density parameter $\Omega_{k0}$, we find that percent-level precision is possible using only siren and galaxy data in $z\le0.5$. Given the limited available volume observed at lower redshifts, it is crucial to try to explore the data there to the fullest. For $H_0$, in the same redshift range the inclusion of clustering information considerably improves the results arising from distance measurements only, resulting in 30\% smaller uncertainties. Given the current efforts to better understand the reasons behind the Hubble tension, this is an important possible gain, and has the advantage of only relying on data in this redshift range. Moreover, sirens should then also outperform the precision on $H_0$ from either galaxy power spectrum measurements at higher redshifts or the CMB by an order of magnitude.  And together with higher redshift measurements, we should be able to perform precise measurements of this parameter at different redshift ranges.

We also find that the inclusion of $10^\circ$ priors on the inclination of the orbital plane from electromagnetic observations have little impact on the clustering performance alone, but do lead to important improvements on the inferred KN distances and thus also on the combined \six\ + distance approach, which fully explores the cosmological information from KN.  As noted before, it remains to be seen if future photometric KN observations alone will be able to achieve similar constraints on the inclination on average.

Our forecasts for $\sigma_{\gamma}$ are less promising than what was found in~\cite{Palmese:2020kxn}, to wit $\sigma_{\gamma}\sim 0.02-0.03$, even though they only assume a \three\ method. This discrepancy seems to be due to the analysis of the EM follow-ups adopted, as our predictions for the capabilities of EM counterpart observations with future telescopes resulted in a much lower number of joint GW+EM measurements.

We remark that in our forecasts we have made a number of conservative assumptions. We have assumed that for each event the EM follow-up search area is 40 deg${}^2$, instead of simply the smaller $\Delta \Omega$, and also required three epochs of follow-up observations for each KN. We also limited the follow-up at any given time to only $40\%$ of the sky. The total accessible area in the sky at any given night is larger, specially if follow-ups could be conducted at lower altitudes. Finally, we restricted ourselves to the Einstein Telescope alone, but the Cosmic Explorer or other concurrent GW facilities may be operational at the same time, which would increase both the amount of events observed and the precision and the precision in the area determination. Together, these assumptions may result in similar constraints being obtained with less than the 5-year observational period here considered. Nevertheless, the still very large uncertainty in the rate of BNS mergers means that the number of years needed to reach our forecast precision could change significantly when this quantity is better understood in the course of the next observational runs of the LVC.

In any case, the \six\ method applied to standard sirens can provide competitive clustering measurements when compared to SN, being better than the configuration labeled as conservative in \cite{Quartin:2021dmr}, especially when it comes to $H_0$. The results are also more sensitive to the number density of BNS, so if we improve further on our follow-up capabilities even more cosmological information can be obtained.

\section*{Acknowledgments}

We thank Valerio Marra, Ribamar Reis, Rogério Rosenfeld and Riccardo Sturani for useful comments. VA was supported by the Brazilian research agency CNPq. MQ is supported by the Brazilian research agencies CNPq, FAPERJ and CAPES. LA acknowledges support from DFG project  456622116. TC is supported by the FARE Miur grant `ClustersXEuclid' R165SBKTMA and the INFN INDARK PD51 grant. This study was financed in part by the Coordenação de Aperfeiçoamento de Pessoal de Nível Superior - Brasil (CAPES) - Finance Code 001. We acknowledge support from the CAPES-DAAD bilateral project  ``Data Analysis and Model Testing in the Era of Precision Cosmology''.

\section*{Data availability}

The data underlying this article will be shared on reasonable request to the corresponding author.

\bibliography{references,scaling_bib}

%%%%%%%%%%%%%%%%% APPENDICES %%%%%%%%%%%%%%%%%%%%%

\appendix

\section{Short review on distance posterior approximations}\label{app:app}

In this appendix we briefly review the Cutler and Flanagan (CF) approximation \cite{Cutler:1994ys} for the posterior distribution on luminosity distance, which was rederived and corrected in \cite{Chassande-Mottin:2019nnz}.

Under the following assumptions that:
\begin{itemize}
    \item the noise is Gaussian, stationary and uncorrelated between detectors,
    \item we work at leading post-Newtonian order,
    \item the sky localisation is known from an EM counterpart,
    \item redshifted chirp mass and time to coalescence are very precisely measured and do not impact the measurement of distance, inclination $\iota$, polarization and phase at coalescence,
\end{itemize}
the posterior distribution on distance and inclination, marginalised over polarization and phase, can be written as:
\begin{equation}
\label{posteriordL}
\begin{aligned}
&p\left(d', v'|\vec{\theta}\right)\propto &  \\
& \;\;\left\{
\begin{array}{ll}
    \!\!\pi\left(d',v'\right)\exp\bigg[-\frac{\rho_{0}^{2}\sigma_{d}}{2} \Big[\left(1-\epsilon_{d}\cos4\overline{\Psi}\right) \left(v-d'^{-1}v'\right)^{2} \\
    \quad +\left(1+\epsilon_{d}\cos4\overline{\Psi}\right) \left(\chi_{+}-d'^{-1}\chi'_{+}\right)^{2}\Big]\bigg], \qquad\qquad \epsilon_{d}\neq 0\\
    \!\!\pi\left(d',v'\right)I_{0}\left(z_{+}\right)I_{0}\left(z_{-}\right) \\
    \quad \exp\left[-\frac{\rho_{0}^{2}\sigma_{d}}{2} \left[\left(\chi^{2}_{+}+v^{2}\right)+d'^{-2} \left(\chi_{+}^{'2}+v'^{2}\right)\right]\right], \; \epsilon_{d}=0
\end{array} \right.
\end{aligned}
\end{equation}
where $\pi\left(d',v'\right)$ is the prior, $d = d_L/d_L^{\rm true}$ with $d_L^{\rm true}$ being the true luminosity distance, $v = \cos \iota$, $\chi_{+}=0.5\left(1+v^{2}\right)$, $z_{\pm} = \frac{\rho_{0}^{2}\sigma_{d}}{2}\left(\chi_{+}\pm v\right)\left(\chi'_{+}\pm v'\right)$, and $\overline{\Psi} = \Psi+ \arctan\left[2\Theta^{+\times}/\Theta^{++}-\Theta^{\times\times}\right]/4$. $I_{0}$ is the modified Bessel function. Unprimed quantities refer to the true values, and primed ones to where the posterior is evaluated at. We made use of the tensor $\Theta^{AB}$, defined for $A, B = +, \times$ over a network of detectors as
\begin{equation}
\Theta^{AB}=\frac{F^{A}_{a}F^{B}_{b}\delta_{ab}\int_{0}^{\infty}f^{-7/3}S_{n,a}^{-1}\left(f\right)df}{\int_{0}^{\infty}f^{-7/3}S^{-1}_{\textrm{n,aver}}\left(f\right)df},
\end{equation}
where $S_{n,a}$ is the power spectral density in detector $a$ and $S^{-1}_{\textrm{n,aver}}$ is the average of the inverse of the power spectral densities over all detectors, in our case the three detectors that make ET. We define
\begin{equation}
    \rho_{0}^{2} = \frac{5G\mathcal{M}^{5/3}}{6\pi^{4/3}c^{3}}\int_{0}^{\infty}df\frac{f^{-7/3}}{S_{\textrm{n,aver}}\left(f\right)}\,,
\end{equation}
which can be seen as the SNR squared for an overhead face-on source in the case of a single detector. From the tensor $\Theta^{AB}$, $\epsilon_d$ and $\sigma_d$ are defined as
\begin{align}
    \sigma_{d} & = 0.5 \,\textrm{Tr}\Theta^{AB},\\
    \epsilon_{d} & = \sqrt{\frac{2\, \textrm{Tr}\left(\Theta^{AB}\right)^{2}}{\left( \textrm{Tr}\Theta^{AB}\right)^{2}}-1}.
\end{align}

\begin{figure}
	\centering
	\subfigure{\includegraphics[width=\columnwidth]{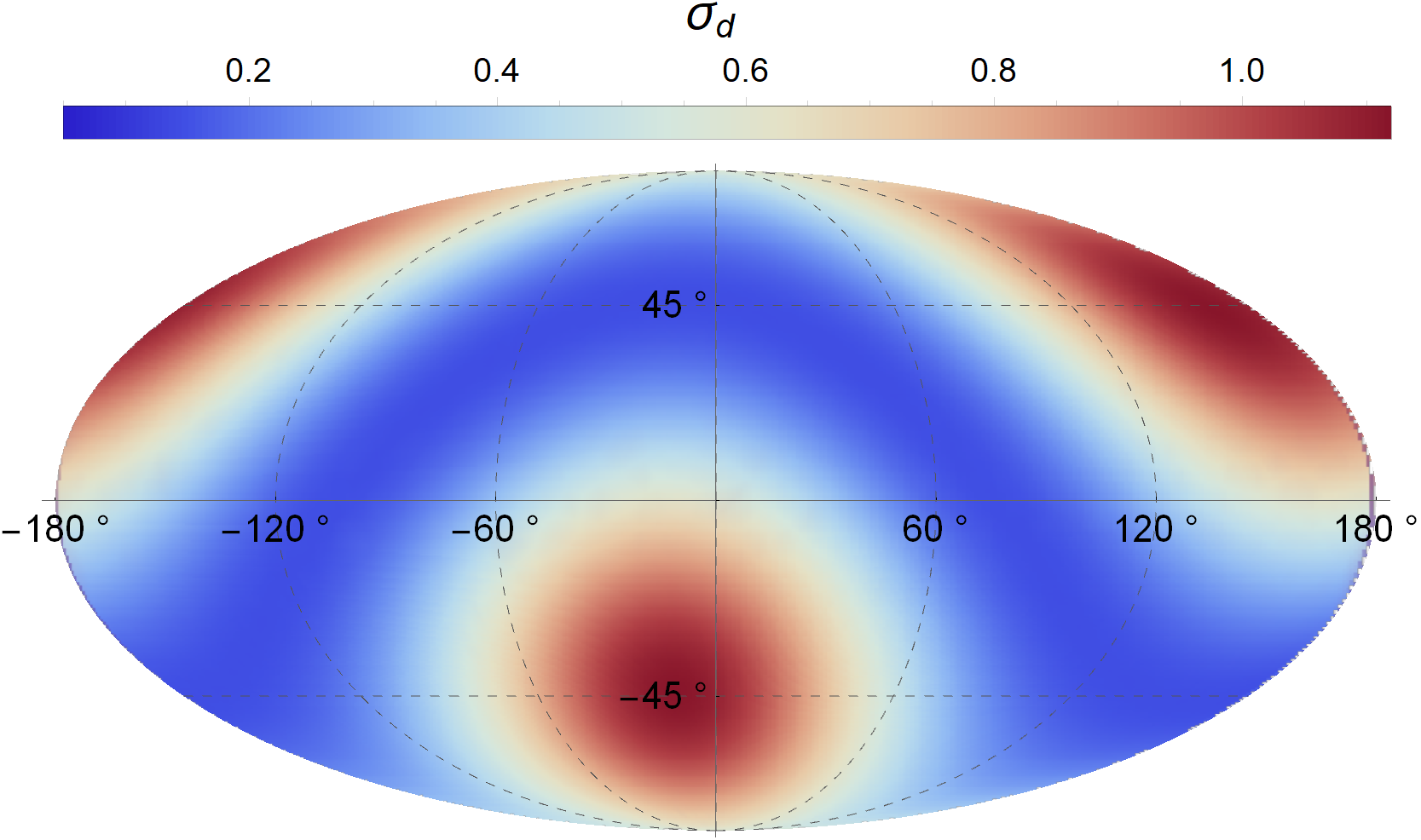}}
	\subfigure{\includegraphics[width=\columnwidth]{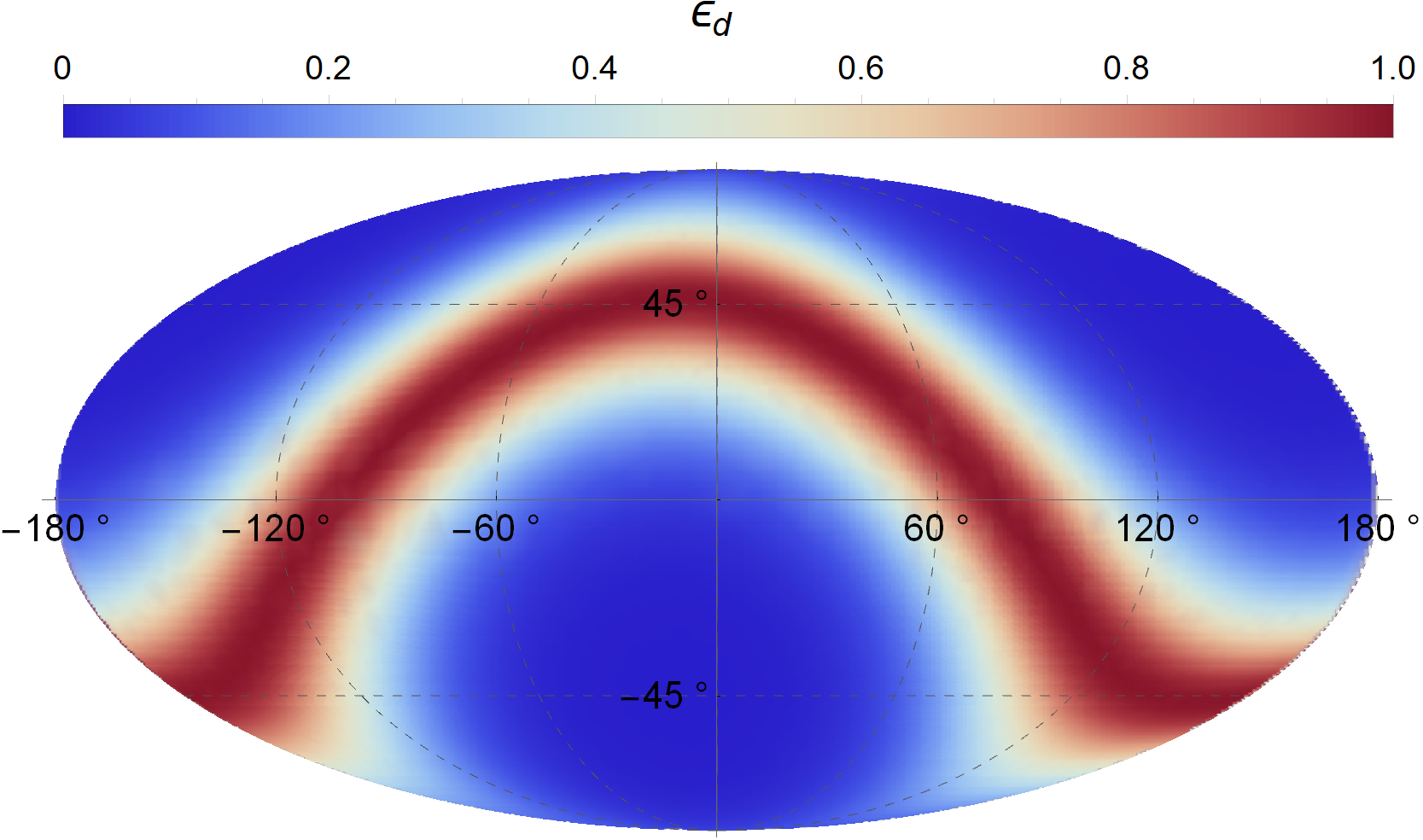}}
    \caption{Amplitude sensitivity $\sigma_d$ and polarization sensitivity $\epsilon_{d}$ over the sky for the ET detectors.}
	\label{ETsensitivity}
\end{figure}

The quantity $\epsilon_{d}$ takes values in the range from 0 to 1 and measures the detector's ability to distinguish the two polarization amplitudes. In the limit $\epsilon_{d}\rightarrow1$ the detector can only estimate one of the polarization tensor components, which makes it difficult to estimate the parameters of the gravitational wave (see discussion in section IV.A of~\cite{Chassande-Mottin:2019nnz}). $\sigma_{d}$ is called the amplitude sensitivity and is related to the loudness of the signal in the network of detectors. large values being favorable. Figure \ref{ETsensitivity} shows how $\sigma_d$ and $\epsilon_d$ changes with the sky position for the ET detectors. Comparing this figure with Figures 1, 2 of~\cite{Chassande-Mottin:2019nnz} we can observe the improvement in the sensitivity achieved by the ET over the detectors currently in operation.

We validate our results produced with the CF approximation by comparing with those found by a Bayesian analysis performed with the Bilby package~\citep{Ashton:2018jfp}. For the Bayesian inference, we use \texttt{emcee} as our sampler, with 100 steps, 200 walkers, and the IMRPhenomPv2 waveform as our signal template. The MCMC was performed varying $\{\mathcal{M}_z, q=m_2/m_1, d_L, \varphi_c, \psi, \iota\}$ and fixing the remaining parameters at their injection values. The analysis shows that the results of the CF approximation are equivalent to the one obtained with MCMC when $\epsilon_{d}\lesssim 0.85$, which is in agreement with what was found in \cite{Chassande-Mottin:2019nnz} ($\epsilon_{d} \lesssim 0.8$). To correctly describe the uncertainty on $d_L$ we imposed this cut in our catalog. The fraction of events with $\epsilon_d\lesssim0.85$ is approximately 0.85 of the BNS mergers, but for the detected events (i.e., with $\rho_{\rm net}>12$) this fraction is 0.98. Thus, the selection cut $\epsilon_{d}$ has an almost negligible impact the number of events in our analysis.

\section{Fraction of overlapping events }\label{app:overlap}

We assessed the impact of overlapping events on the number of observed objects presented in Tables~\ref{tab:texp-comparison} and~\ref{tab:Ftime} by Monte-Carlo sampling a time series of events and assuming that the events are observed serially and that kilonovae are detectable during a total of $\delta\,t$ hours after the event. The fraction of overlapping events are depicted in Table~\ref{tab:overlap}. The overlapping fraction variance was estimated to generate 100 random time series for each configuration. The number of expected kilonovae would be only strongly affected in all surveys if kilonovae happened to be detectable during only a fraction of a day. Only ZTF would still be significantly affected if kilonovae lasted only for 24 hours.

In our simulations we did not find any scenario where an event could not be observed if kilonovae lasted at least two days. The detectability and best strategy to observe kilonovae have been already studied~\citep[see, for instance,][]{Cowperthwaite:2015kya, Mochkovitch:2021prz, Chase:2021ood}. The exact details for the best observation strategy go beyond the scope of our work; still, we can neglect the impact of overlapping events for two reasons. Due to the slow decaying of kilonovae in certain magnitudes, the detectability of a large fraction of the sample will be larger than two days. Secondly, our simulation assumed that the observation strategy followed the events serially; any strategy optimization will reduce the overlapping fraction presented in Table~\ref{tab:overlap}.

\begin{table}
    \centering
    \begin{tabular}{c|c|c|c}
    Survey & \multicolumn{3}{c}{Overlapping Fraction} \\
    & $\delta\,t=12$ h & $\delta\,t=24$ h & $\delta\,t=48$ h \\ \hline
    LSST & $16.1 \pm 0.5 \%$ & 0 & 0 \\
    Mephisto & $36.1 \pm 1.2 \%$ & $3.8 \pm 0.5 \%$ & 0 \\
    WFST & $30.6 \pm 1.1 \%$ & $1.4 \pm 0.3 \%$ & 0 \\
    ZTF & $69.6 \pm 1.5 \%$  & $63.7 \pm 1.4 \%$ & 0 \\ \hline
    \end{tabular}
    \caption{The fraction of overlapping events estimated generating $100$ random time-series for each configuration assuming kilonovae are detectable $\delta\,t$ hours after the event.
    }
    \label{tab:overlap}
\end{table}

\section{Dependence on the distance measurement uncertainties}\label{app:sint}

The inclusion of information on the inclination of the BNS orbit, which can be independently obtained from EM observations, is capable of improving the precision of the sirens distance measurements. A $10^\circ$ constraint on this inclination leads to improvements on the relative distance of a given source by between 15--25\%. Nevertheless as discussed in Section 4, this leads only to very minor improvements on the clustering results from BNS. This is also in opposition to what was found for the case of supernovae by~\cite{Amendola:2019lvy}, where a much stronger dependency of the \three\ and \six\ methods on $\sigma_{\rm int}$ was found. And neglecting the small extra lensing scatter, $\sigma_{\rm int} = (5/\ln 10) \sigma d_L / d_L$.

\begin{figure}
    \centering
    \includegraphics[width=\columnwidth]{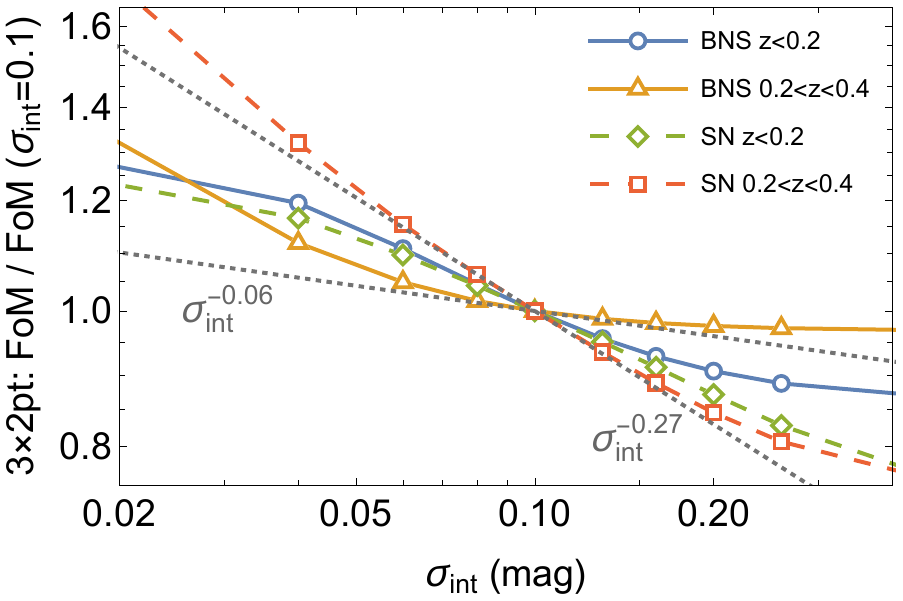}
    \caption{Dependency of the \three\ method on the standard candle distance uncertainty. The Figure of Merit (FoM) used is the inverse area of the ellipses in the $\sigma_8-\gamma$ plane. Solid (dashed) lines are the results for our BNS (SN) forecasts. Also shown are dotted gray lines with example power laws that roughly fit the BNS and SN curves for $0.2<z<0.4$. As can be seen, the low number density of BNS means that the influence of $P_{vv}$ is smaller and thus its contribution to \three\ and \six\ are also smaller, which explains the lower sensitivity to $\sigma_{\rm int}$. The uncertainties due to $P_{vv}$ alone scale instead roughly as $\sigma_{\rm int}^{-2}$. }
    \label{fig:sint}
\end{figure}

The reason for this small effect is basically the low expected number density of BNS, when compared to that of LSST supernovae. To understand this, we remark first that the uncertainties due to $P_{\rm vv}$ alone scale instead roughly as $\sigma_{\rm int}^{-2}$. However, the weight of the velocities on the full \three\ and \six\ methods depend on the number density of velocity tracers. The low number density of BNS means that the influence of $P_{\rm vv}$ is small and thus its contribution to \three\ and \six\ are also smaller, which explains the lower sensitivity to $\sigma_{\rm int}$. Figure~\ref{fig:sint} illustrates this, and compares the inverse area of the ellipses in the $\sigma_8-\gamma$ plane, which is our Figure of Merit (FoM), as a function of $\sigma_{\rm int}$, assumed constant and the same for all sources (contrary to our baseline KN analysis in Section~\ref{sec:forecasts}). For illustration purposes we normalize all results to the one obtained for $\sigma_{\rm int} = 0.1$ mag. For very low redshifts both SN and KN clustering exhibit small dependencies on $\sigma_{\rm int}$, but for $0.2<z<0.4$ the SN Figure of Merit scale as $\sigma_{\rm int}^{-0.27}$, whereas the KN clustering one only as $\sigma_{\rm int}^{-0.06}$.

\label{lastpage}

\end{document}